\documentclass[aps,prl,10pt,twocolumn,superscriptaddress,nofootinbib]{revtex4-2}
\usepackage[T1]{fontenc}
\usepackage{amsmath,amssymb,bm,xcolor}
\usepackage{graphicx,needspace}
\usepackage[colorlinks,linkcolor=blue,citecolor=blue,urlcolor=blue]{hyperref}
\newcommand{\thorn}{\mbox{\fontencoding{T1}\selectfont\th}}
\newcommand{\vO}{\bm{\Omega}}
\newcommand{\vS}{\bm{S}}
\newcommand{\Dr}{\mathcal{D}}

\allowdisplaybreaks
\begin{document}

\title{Separation of variables for massive spin-2 fields on the Kerr black hole}

\author{Oscar J. C. Dias}
\affiliation{STAG research centre \& Mathematical Sciences, University of Southampton, Highfield Campus, UK}
\author{Jorge E. Santos}
\affiliation{DAMTP, Centre for Mathematical Sciences, University of Cambridge, Wilberforce Road, Cambridge CB3 0WA, UK}

\date{\today}

\begin{abstract}
Massive spin-2 fields on the Kerr black hole generally obey coupled partial differential equations. We show that these equations separate through two commuting angular operators acting on five components of the field, its curl and its linearised Weyl tensor. A joint angular eigenfunction fixes the relative profiles of the five polarisations and leaves one second-order radial equation. The separation constants satisfy an algebraic relation that follows from the angular operators themselves. The construction removes the obstruction to a Teukolsky-type equation by a radial transport identity, then eliminates the additional radial solution through a first integral. It provides ordinary differential equations for massive spin-2 fields on Kerr, and, equivalently, for Kaluza–Klein perturbations of the Kerr black string.
\end{abstract}
\maketitle

\paragraph{Introduction.}
The Teukolsky equations reduce massless perturbations of the Kerr black hole to ordinary differential equations (ODEs)~\cite{Teukolsky:1972my,Teukolsky:1973ha}. This makes it possible to compute quasinormal modes and superradiant scattering through a radial boundary-value problem and an angular eigenvalue problem. The underlying geometry also admits a hidden symmetry: Carter's constant~\cite{Carter:1968rr} is generated by a Killing tensor, which belongs to the hierarchy of symmetries associated with the principal closed conformal Killing-Yano tensor~\cite{Frolov:2017kze}.

For massive fields, separation is more difficult. For a massive spin-1 field, an ansatz based on the principal tensor gives separated equations for the Proca field~\cite{Lunin:2017drx,Frolov:2018ezx}.
Until now, however, separation of the massive spin-2 equations had remained elusive. The field has five polarisations whose radial and angular dependence is coupled. Its superradiant spectrum at generic Kerr rotation has nevertheless been obtained by solving the coupled partial differential equations (PDEs) in $(r,\theta)$~\cite{Dias:2023ynv}. The same field also contains the spherically symmetric instability of Schwarzschild in massive gravity~\cite{Babichev:2013una,Brito:2013wya,Dias:2023ynv}, related to the Gregory--Laflamme instability of a black string~\cite{Gregory:1993vy}.

Nevertheless, the coupling between polarisations need not prevent separation. The coupled gravito-electromagnetic equations of Kerr--Newman separate through a matrix integrating factor~\cite{Hintz:2026jiw}. For singly spinning Myers--Perry black holes, commuting matrix angular operators organise a coupled gravitational sector into modes with separated angular and radial dependence~\cite{Santos:2026pbu}. This suggests looking for an angular profile with several components, rather than a separate scalar equation for each polarisation, in the massive spin-2 Kerr system as well.

In this Letter we show that the Fierz--Pauli equations for a massive spin-2 field on Kerr ~\cite{Fierz:1939ix,Hinterbichler:2011tt,deRham:2014zqa} separate in this way. We construct two commuting angular operators, identify their covariant symmetry operators, and obtain a single second-order radial equation for each joint angular eigenfunction. The Supplemental Material (SM)
gives the transport identities, the spectral polynomial and the reconstruction argument, together with detailed derivations of the key relations used below.
We then use the separated equations to substantially extend the numerical results available from coupled PDEs~\cite{Dias:2023ynv}. We reproduce previously computed dipole frequencies, resolve the $\ell=m=2,3$ superradiant branches, and map their competition with the Gregory--Laflamme instability, including close to extremality. The reduction to ODEs makes these calculations substantially faster and provides an efficient framework for exploring the wider quasibound-state and quasinormal spectra and their implications for observational signatures of massive gravitons.

\paragraph{Setup.}
A massive spin-2 field $h_{ab}$ with mass $\mu\neq0$ on the Ricci-flat Kerr background $ds^2_{\hbox{\tiny Kerr}}$~\cite{Kerr:1963ud} obeys~\cite{Fierz:1939ix,Hinterbichler:2011tt}
\begin{equation}\label{eq:FP}
\delta R_{ab}[h]+\tfrac12\mu^2h_{ab}=0,\qquad \nabla^bh_{ab}=0,\qquad h^a{}_a=0,
\end{equation}
where $\delta R_{ab}[h]$ is the linearised Ricci tensor.
The divergence and trace constraints in \eqref{eq:FP} leave five physical polarisations. Equivalently, $H=h_{ab}e^{{\rm i}\mu z}{\rm d}x^a {\rm d}x^b$, with $H_{zA}=0$ (with $A$ running over all five spacetime coordinates), is a vacuum perturbation of the five-dimensional Kerr string ${\rm d}s^2_{\hbox{\tiny Kerr}}+{\rm d}z^2$. The mass $\mu$ is then its wave number along the string. As for the Proca field, the four-dimensional massive spin-2 field has no residual gauge freedom.

We use Boyer--Lindquist coordinates $x^a=\{t,r,\theta,\phi\}$, the Kinnersley tetrad $(l,n,m,\bar m)$ with $l\cdot n=-1$ and $m\cdot\bar m=1$~\cite{Teukolsky:1972my,Teukolsky:1973ha}, and modes proportional to $e^{-{\rm i}\omega t+{\rm i}m\phi}$, where $\omega$ is the frequency and $m$ the azimuthal number. 
We also use the Geroch--Held--Penrose (GHP) formalism~\cite{Geroch:1973am}, with derivative operators $\{\thorn,\thorn',\eth,\eth'\}$ and spin coefficients such as $\rho$ and $\tau$, defined in SM Sec.~\ref{sm:s:tools}. GHP scalars carry boost weight $b$ and spin weight $s$. On Kerr, we define 
\begin{gather}
\zeta=r-{\rm i}a\cos\theta,\quad \Sigma=\zeta\bar\zeta,\quad
\Delta=r^2-2Mr+a^2,\nonumber\\
K=(r^2+a^2)\omega-am,\qquad D=\partial_r-{\rm i}K/\Delta \nonumber\\
\rho=-1/\zeta\, \qquad \Psi_2=-M/\zeta^3\,,\label{eq:defs}
\end{gather}
where $M$ and $a$ are the Kerr mass and rotation parameters~\cite{Kerr:1963ud,Teukolsky:1972my,Teukolsky:1973ha}.

The five field and curvature variables we use to separate the Fierz--Pauli system form the column vector
\begin{equation}\label{eq:Omega}
\vO=(2\Psi_0,\sqrt2 X_1,h_{ll},\sqrt2\bar X_1,2\bar\Psi_0)^{\mathsf T}.
\end{equation}
The entries have GHP boost weight two and spin weights $s=2,1,0,-1,-2$, respectively. They consist of the field component $h_{ll}$ and components of its curl and linearised Weyl tensor, defined by  
\begin{align}\label{eq:field-components}
& X_1=F_{lml},  \:\:
\Psi_0=\delta C_{lmlm}, \:\:
\bar X_1=F_{l\bar m l}, \:\:
\bar\Psi_0=\delta C_{l\bar m l\bar m}, \nonumber \\
& \hbox{where}\quad F_{abc}\equiv\nabla_a h_{bc}-\nabla_b h_{ac}
\end{align}%
is the curl of $h$ and $C$ is the Weyl tensor.
These components vanish on the Kerr background, so they are unaffected by infinitesimal tetrad rotations. The bar denotes the $m\leftrightarrow\bar m$ partner, rather than complex conjugation of a Fourier mode. 
The five components of $\vO$ encode the five polarisations collectively; they are not individually decoupled polarisation amplitudes.

\paragraph{Angular separation.}
Our aim is to separate the angular dependence from the radial dynamics while keeping the coupling between the five curvature components in $\vO$.
Recall that for a scalar field, an angular harmonic fixes the angular dependence of a single amplitude.
Here, the harmonic $\vS(\theta)$ has five components and fixes the relative angular profiles of $(\Psi_0,X_1,h_{ll},\bar X_1,\bar\Psi_0)$. 

The key objects are two $5\times5$ angular matrix operators, a first-order operator $K_1$ and a second-order operator $K_2$, given explicitly in SM Sec.~\ref{sm:operators}. Both are independent of $r$ and $M$, and satisfy the exact operator identities
\begin{equation}\label{eq:FK}
[K_1,K_2]=0,\qquad F(K_1,K_2)=0.
\end{equation}
The polynomial $F$, given in SM Eq.~\eqref{sm:eq:Fexpl}, has degree five in its first argument and degree two in its second.
These identities organise the angular
problem into joint eigenfunctions of $K_1$ and $K_2$:
\begin{equation}\label{eq:sep}
\vO=e^{-{\rm i}\omega t+{\rm i}m\phi}f(r)\,\vS(\theta),\qquad K_1 \vS=\lambda \vS,\quad K_2 \vS=\nu \vS.
\end{equation}
Their eigenvalues $\lambda$ and $\nu$ must obey
\begin{equation}\label{eq:F}
F(\lambda,\nu;a,m,\omega,\mu)=0,
\end{equation}
so $\lambda$ and $\nu$ cannot be chosen independently at fixed $(a,m,\omega,\mu)$. Together with regularity at $\theta=0,\pi$, this relation selects the allowed five-component harmonics $\vS(\theta)$ and their eigenvalues.

The angular operators have a covariant origin. The principal closed conformal Killing-Yano tensor generates a symmetry $\mathcal K_1$ of the Fierz--Pauli equations, while a completion of Carter's operator gives a second symmetry $\mathcal K_2$ (see Sec.~\ref{sm:operators} of SM). On solutions,
\begin{equation}\label{eq:intertwine}
\vO[\mathcal K_ih]=K_i\,\vO[h]
\end{equation}
so the angular operators are the actions of these spacetime symmetries on $\vO$. Their covariant expressions and the derivation of Eq.~\eqref{eq:intertwine} are given in SM Secs.~\ref{sm:operators} and~\ref{sm:s:sym}. $K_1$ contains explicit factors of $\mu^{-2}$, so this construction assumes $\mu\neq0$; the massless limit requires a separate treatment, as is standard in the Fierz--Pauli system ~\cite{Hinterbichler:2011tt}.

Once the joint angular eigenfunction $\vS(\theta)$ is fixed, the remaining problem is to determine the common radial amplitude $f(r)$ in \eqref{eq:sep}.

\paragraph{Radial transport.}
A central result is that the field equations \eqref{eq:FP} imply the closed identity (see SM Sec.~\ref{sm:s:transport-Statement})
\begin{equation}\label{eq:transport}
\mathcal E[\vO]\equiv 
\thorn \big[(\Dr+K_2)\vO\big]+\big(2\mu^2K_1-4\mu^2r+8{\rm i}\omega\big)\vO=0,
\end{equation}
where $\thorn$ reduces to $D$ defined in \eqref{eq:defs} on the quantities of this Letter,
and $\Dr$ is the spin-2 Teukolsky radial operator with a mass term in SM Eq.~\eqref{eq:D}. Equation~\eqref{eq:transport} is third order in $r$, while its angular dependence enters only through $K_1$ and $K_2$.

To obtain \eqref{eq:transport}, consider first its $s=2$ component. Teukolsky's identity for $\Psi_0$ has a source containing a term proportional to $F_{lmm}$. This component is absent from $\vO$ in \eqref{eq:Omega} and prevents the second-order system from closing. In the Kerr-string interpretation, it corresponds to the obstruction associated with the expansion of the principal null congruence~\cite{Durkee:2010qu}.
The obstruction can be removed by taking a $\thorn$-derivative and using the cyclic identity of the curl,
\begin{equation}\label{eq:Bianchi}
(\thorn-\bar\rho)F_{lmm}=(\eth-\bar\tau')X_1-2\Psi_0.
\end{equation}
Together with $\thorn\bar\zeta=1$ and $\bar\rho\bar\zeta=-1$, this gives
$\thorn(\bar\zeta F_{lmm})=\bar\zeta[(\eth-\bar\tau')X_1-2\Psi_0]$.
Thus the differentiated equation depends only on $X_1$ and $\Psi_0$, with no $F_{lmm}$, and gives the $s=2$ row of \eqref{eq:transport}. The remaining rows of~\eqref{eq:transport} close through analogous cyclic and divergence identities. The full derivation is given in SM Sec.~\ref{sm:s:transport}.

\paragraph{The physical radial equation.}
The five angular components remain coupled, but the ansatz \eqref{eq:sep} separates their angular profiles from the common radial amplitude $f(r)$. On a regular joint angular eigenfunction $\vS(\theta)$, the transport identity~\eqref{eq:transport} reduces to
\begin{equation}\label{eq:L3}
L_3f\equiv D\big[(\Dr+\nu)f\big]+\big(2\mu^2\lambda-4\mu^2r+8{\rm i}\omega\big)f=0 .
\end{equation}
For an ansatz $f\sim e^{\kappa r}r^p$, \eqref{eq:L3} at large $r$ requires
\begin{equation}
(\kappa-{\rm i}\omega)(\kappa^2-\mu^2+\omega^2)=0.
\end{equation}
Thus, the two massive branches have $\kappa=\pm\sqrt{\mu^2-\omega^2}$. 
The third branch has massless asymptotics, $\kappa={\rm i}\omega$, and is introduced by the extra radial differentiation that was performed to get \eqref{eq:transport}.
A physical Fierz--Pauli solution must also satisfy a first integral that removes this
unphysical solution (see SM Eq.~\eqref{sm:eq:L2op} and Sec.~\ref{sm:s:recon}).

Let $A(r)=\sum_{k=0}^{4}A_k r^k$ be the quartic polynomial defined in SM Sec.~\ref{sm:s:radial}, normalised by $A_4=\mu^4$. Its coefficients $A_k$ follow from an adjoint equation, and the polynomial terminates precisely when the eigenvalue condition~\eqref{eq:F} holds. Define
\begin{equation}\label{eq:radial}
L_2f\equiv A(\Dr+\nu)f-\Delta A'f'\\
+\big[\Delta A''-(2\Delta'+{\rm i}K)A'\big]f=0.
\end{equation}
The operator identity $D\circ L_2=A L_3$ then gives $L_2f=C\exp({\rm i}\!\int K/\Delta\,dr)$ for a solution of $L_3f=0$. The Fierz--Pauli equations enforce $C=0$. This follows locally from the reconstruction identities in SM Sec.~\ref{sm:s:recon}; the large-$r$ behaviour gives the same conclusion, since the massive exponential dependence of $L_2f$ is incompatible with $C\exp({\rm i}\!\int K/\Delta\,dr)$ unless $C=0$.
Equation~\eqref{eq:radial} is therefore the second-order radial equation for the physical amplitude $f(r)$. Multiplying it by $\Delta$ gives polynomial coefficients. For generic parameters, the zeros of $A$ are apparent singularities and the remaining singular points are the two horizons of Kerr and infinity. 

For fixed $(\omega,m)$, one first solves the regular joint angular eigenvalue problem~\eqref{eq:sep} and then the radial equation~\eqref{eq:radial}. Quasibound states require an ingoing solution at the future horizon and decay at infinity; quasinormal modes instead require outgoing behaviour at infinity. The local reconstruction operator $\mathcal Q$, given in SM Eq.~\eqref{sm:eq:Qrec}, recovers the metric perturbation $h$ from $\vO$ and establishes the converse: solutions of the physical separated system reconstruct Fierz--Pauli fields (see SM Sec.~\ref{sm:s:recon}). Angular completeness gives a convergent expansion in separated solutions, with finite Jordan-chain combinations included when necessary at complex frequencies (SM Sec.~\ref{sm:s:recon}).
$\mathcal Q$ has differential order at most six. Axis regularity is preserved, and for ingoing $f$ the reconstructed field is regular on the future horizon except possibly at $\omega-m\Omega_H\in\{0,\pm {\rm i}\kappa_+,2{\rm i}\kappa_+\}$, where $\kappa_+$ is the surface gravity and $\Omega_H$ is the horizon angular velocity (SM Sec.~\ref{sm:s:recon}).
% Exceptional horizon frequencies: SM Sec.~\ref{sm:s:recon}.
% Reconstruction coefficients: ancillary file anc/Q_reconstruction.wl.

\paragraph{Schwarzschild limit.}
For $a=0$, the angular couplings in $K_2$ vanish and $\nu=-\ell(\ell+1)$ , the familiar eigenvalue of spherical harmonics. The spectral polynomial factorises into quadratic and cubic factors, corresponding to the axial and polar sectors (SM Sec.~\ref{sm:s:schw}). These are the massive counterparts of the axial and polar sectors of massless gravitational perturbations~\cite{Regge:1957td,Zerilli:1970se,Kodama:2003jz}. For $\ell=1$, only the components with $|s|\leq1$ remain. Their separation constants are $\lambda=-{\rm i}\omega/\mu^2$ in the axial sector and the two roots of $\lambda^2+{\rm i}\omega\lambda/\mu^2+6/\mu^2=0$ in the polar sector. The latter contains the dipole mode that becomes superradiantly unstable on Kerr. The $\ell=0$ mode has $\lambda=\nu=0$ and contains the four-dimensional counterpart of the five-dimensional Gregory--Laflamme instability, whose rotating continuation persists for $a\neq0$. These limits identify the branches to follow as rotation is increased.

\begin{figure}[!tbp]
\includegraphics[width=\columnwidth]{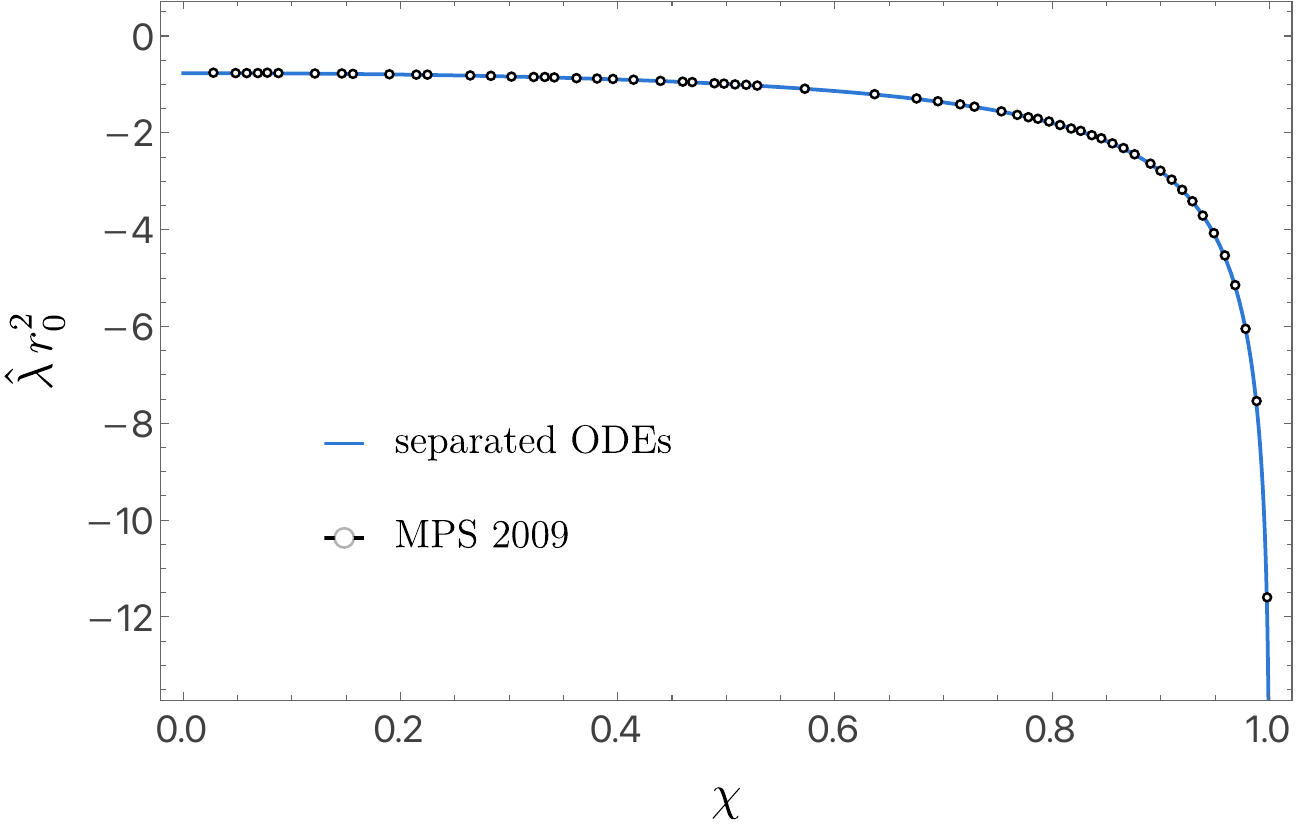}
\caption{The Gross--Perry--Yaffe negative mode of the Kerr instanton, $\hat{\lambda} r_0^2$, with $r_0=2M$, obtained from the separated ODEs at the static threshold, $\hat{\lambda}=-\mu_{\rm GL}^2$ (solid line), compared with the 2D spectral PDE results of Ref.~\cite{Monteiro:2009ke} (circles).}
\label{fig:negative-mode}
\end{figure}

\paragraph{Results.}
We first test the separated ODEs~\eqref{eq:sep} and~\eqref{eq:radial} against existing PDE calculations. Our numerical calculations combine pseudospectral methods, following the schemes of Refs.~\cite{Monteiro:2009ke,Dias:2009iu,Dias:2010eu,Dias:2010maa,Dias:2010gk} and the review~\cite{Dias:2015nua}, with radial integration and matching.
At the stationary axisymmetric threshold, $\omega=m=0$, the Gregory–Laflamme mode determines the critical mass $\mu_{\rm GL}$ and hence the negative eigenvalue $\hat{\lambda}=-\mu_{\rm GL}^2$ of the Kerr instanton. Fig.~\ref{fig:negative-mode} compares this eigenvalue with the two-dimensional spectral calculation of Ref.~\cite{Monteiro:2009ke}. The separated equations reproduce the negative mode across the plotted range of rotation, including its increasingly negative eigenvalue as extremality is approached.

\begin{figure}[!tbp]
\includegraphics[width=\columnwidth]{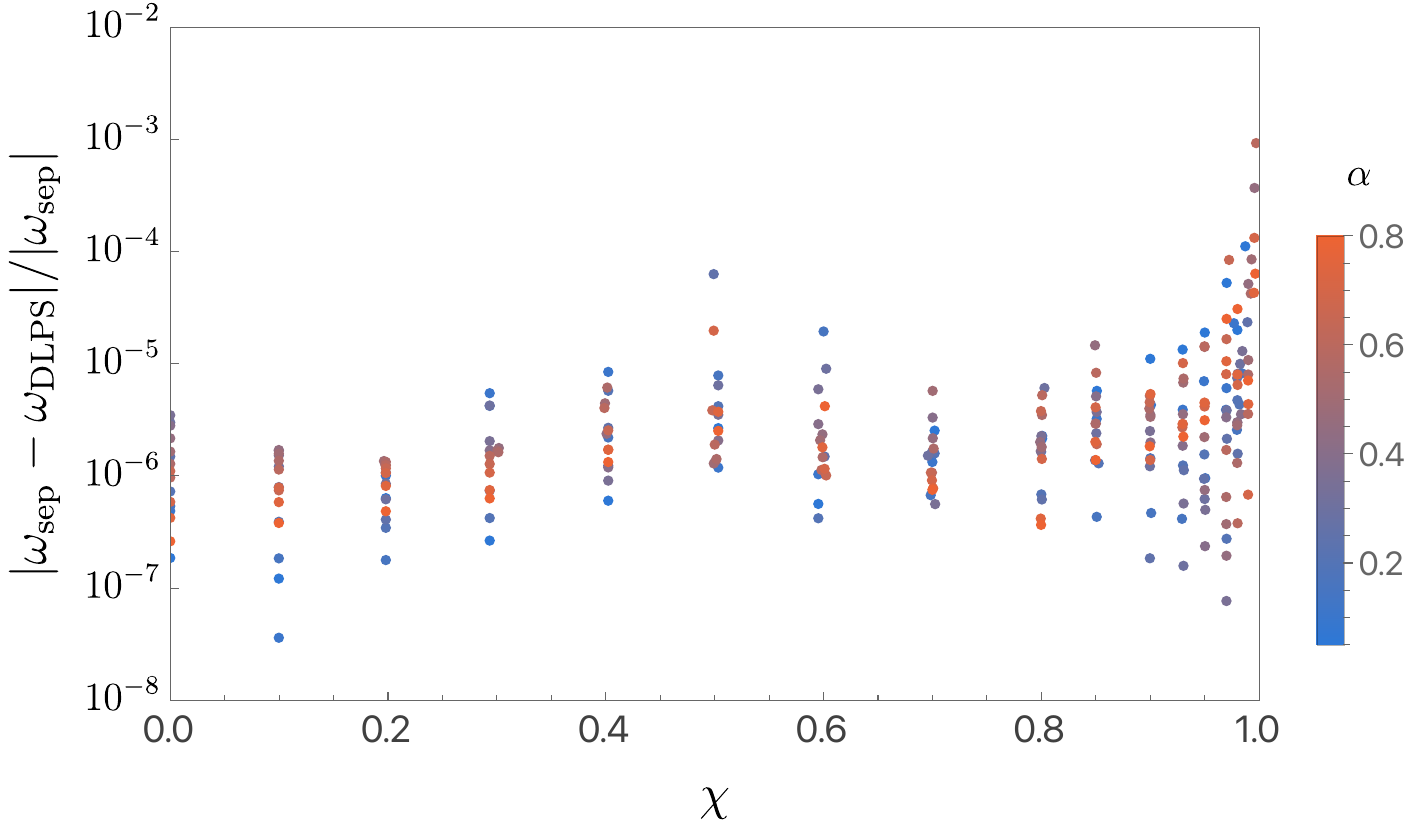}
\caption{Relative difference $|\omega_{\rm sep}-\omega_{\rm DLPS}|/|\omega_{\rm sep}|$ as a function of $\chi=J/M^2$ for a sample of 261 modes from Ref.~\cite{Dias:2023ynv}, coloured by $\alpha=\mu M$.}
\label{fig:recovering}
\end{figure}

We next compare our complex quasibound-state frequencies $\omega_{\rm sep}$ with a sample of 261 frequencies $\omega_{\rm DLPS}$ computed using coupled PDEs in Ref.~\cite{Dias:2023ynv}.
Fig.~\ref{fig:recovering} shows the relative difference between the two calculations as a function of the black-hole spin $\chi=J/M^2$, with colour indicating the mass coupling $\alpha=\mu M$. The calculations agree throughout the sampled parameter range, including rapidly rotating black holes. Together, these comparisons provide independent numerical checks of our separated equations in both the stationary axisymmetric and dynamical non-axisymmetric sectors.

Solving the separated ODEs is substantially faster than solving the coupled PDEs and allows us to resolve much smaller growth rates. In our earlier PDE implementation~\cite{Dias:2023ynv}, numerical errors limited a systematic study of the higher-$m$ instabilities, particularly in the small-spin regime relevant for comparison with perturbative expansions. The separated equations now allow us to follow the $\ell=m=2,3$ branches and determine their competition with the dipole and Gregory--Laflamme modes. The hierarchy of growth rates is illustrated in SM Fig.~\ref{sm:fig:higher-m}.

\begin{figure}[!tbp]
\includegraphics[width=\columnwidth]{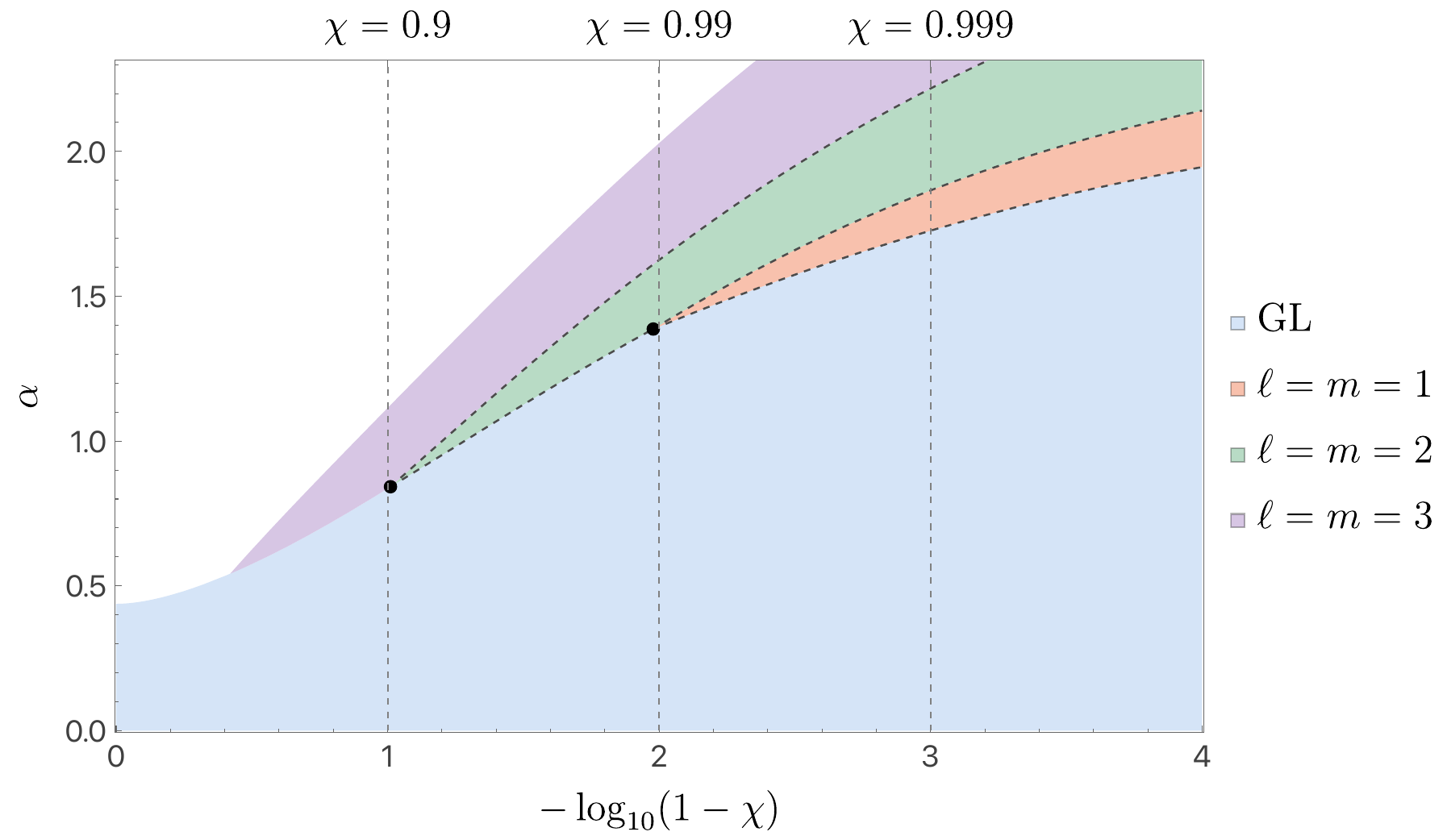}
\caption{Fastest instability in the $(\chi,\alpha)$ plane, with $\chi=J/M^2$ and $\alpha=\mu M$, among the $m=0$ Gregory--Laflamme mode and the $\ell=m=1,2,3$ superradiant modes.  Dots: points where three rates coincide. The boundary of the $\ell=m=3$ region does not test stability against branches with $m>3$.}
\label{fig:growth}
\end{figure}

Finally, we compare the growth rates $\Gamma={\rm Im}\,\omega$ of the axisymmetric Gregory--Laflamme mode and the superradiant branches labelled by $\ell=m=1,2,3$. Fig.~\ref{fig:growth} identifies the fastest-growing mode among these branches in the $(\chi,\alpha)$ plane. The Gregory--Laflamme mode dominates at small $\alpha$. At larger $\alpha$, the dominant superradiant branch depends on both the mass and the spin. The $\ell=m=3$ branch already dominates in part of the diagram at $\chi\gtrsim0.62$, while the $\ell=m=2$ and dipole dominance regions first appear at $\chi\simeq0.90$ and $\chi\simeq0.9895$, respectively. Internal boundaries between coloured regions separate modes with equal growth rates; the two marked disks locate coincidences of three rates. The upper boundary of the $\ell=m=3$ region does not establish stability against all perturbations, since branches with $m>3$ were not computed. The individual frequency surfaces as functions of mass and spin, and the corresponding instability timescales, are presented in SM Sec.~\ref{sm:s:timescales}.

Unlike the six-dimensional Myers--Perry string~\cite{Dias:2022mde}, where a marginal-binding edge also bounds the superradiantly unstable region, the Kerr branches considered here have no such additional boundary: at fixed nonzero spin, unstable bound states persist towards $\alpha\to0$. We discuss this comparison and its physical interpretation in SM Sec.~\ref{sm:s:timescales}.

\paragraph{Discussion.}
The five polarisations of a massive spin-2 field on Kerr can be organised into joint angular eigenfunctions $\bm{S}(\theta)$, which solve the ODE system \eqref{eq:sep}, and evolved through a single second-order radial ODE \eqref{eq:radial} for $f(r)$. The angular profiles remain coupled, but the radial and angular problems separate. Two structures make this possible: a transport identity~\eqref{eq:transport} closes the equations after one radial derivative, and a first integral, given in SM Eq.~\eqref{sm:eq:L2op}, restricts the resulting system to its physical solutions.

This gives a practical route to computing the superradiant and quasinormal spectra of massive spin-2 fields using ordinary differential equations. Through the Kerr-string interpretation, it also gives the radial problem for the rotating Gregory--Laflamme instability. Whether the same transport and symmetry construction extends to Kerr-(A)dS or more general higher-dimensional rotating backgrounds remains an open question~\cite{Frolov:2017kze,Santos:2026pbu}.

The separated formulation has allowed us to study higher-$m$ superradiant modes, including close to extremality, substantially extending the results accessible with the coupled PDE approach of Ref.~\cite{Dias:2023ynv}. Following the strategy of Refs.~\cite{Cardoso:2018tly,Dias:2023ynv}, these results open the way to a more comprehensive study of massive-graviton observational signatures over a wider range of masses and black-hole spins.

The instability thresholds also provide a starting point for constructing new nonlinear solutions. In six-dimensional rotating strings, Gregory--Laflamme and superradiant onsets lead to nonuniform, black resonator and helical strings~\cite{Dias:2022mde,Dias:2022str,Dias:2023nbj}. The thresholds found here motivate analogous constructions for the five-dimensional Kerr string, now including higher-$m$ superradiant branches. In four-dimensional theories whose linearised equations reduce to the Fierz--Pauli system~\eqref{eq:FP}~\cite{deRham:2010kj,Hassan:2011hr,Hassan:2011zd,Babichev:2016bxi,PhysRevD.32.379,PhysRevD.16.953,Hinterbichler:2011tt,deRham:2014zqa}, the marginal modes similarly motivate a search for rotating black holes with massive spin-2 hair. Together, the separated equations and metric reconstruction provide a systematic framework for pursuing these constructions. Determining the nonlinear branches and their thermodynamic properties is a natural next step.

%%%%%%%%%
\paragraph*{Acknowledgements}
Claude (Anthropic) and Codex (OpenAI) were used under the authors’ direction to assist with checking the derivations and computations. J. E. S. thanks OpenAI for providing access to ChatGPT through the ChatGPT for Academic Researchers programme. OJCD acknowledges financial support from the STFC consolidated grant ST/X000583/1. J. E. S. acknowledges partial support from STFC consolidated grant UKRI4224 and from Hughes Hall, Cambridge.

\bibliographystyle{apsrev4-2}
\bibliography{biblio}

\onecolumngrid
\clearpage
\setcounter{equation}{0}
\setcounter{section}{0}
\setcounter{subsection}{0}
\renewcommand{\theequation}{S\arabic{equation}} \setcounter{figure}{0}\renewcommand{\thefigure}{S\arabic{figure}}\renewcommand{\theHfigure}{supp.\arabic{figure}}
\renewcommand{\thesection}{S\arabic{section}}
\renewcommand{\theHequation}{supp.\arabic{equation}}
\renewcommand{\theHsection}{supp.\arabic{section}}
\setcounter{secnumdepth}{3}
\begin{center}
\textbf{\large Supplemental Material}\\[1ex]
\textbf{Separation of variables for massive spin-2 fields on the Kerr black hole}
\end{center}

\noindent  This Supplemental Material presents the instability timescales in Sec.~\ref{sm:s:timescales} and gives the four-dimensional derivation of the separated equations. Section~\ref{sm:operators} lists the coordinate operators and their covariant counterparts. Section~\ref{sm:s:tools} fixes conventions and lists the tools. Section~\ref{sm:s:transport} derives the transport identity row by row. Section~\ref{sm:s:sym} gives the hidden symmetries and the angular operators. Section~\ref{sm:s:radial} reduces the problem to the radial equation. Section~\ref{sm:s:recon} shows that $h\mapsto\vO$ is a bijection onto the solutions of the separated equations, with an explicit inverse. Section~\ref{sm:s:schw} treats the Schwarzschild limit. Equation numbers without an S refer to the Letter; all cross-references are internal to this document.

 The gravitational field of the Kerr black hole background is
\begin{equation}\label{metric}
\mathrm{d}s^2_{\hbox{\tiny Kerr}}=-\frac{\Delta}{\Sigma}\left[\mathrm{d}t-a\sin^2\theta\,\mathrm{d}\phi \right]^2
+\frac{\Sigma}{\Delta}\,\mathrm{d}r^2+\Sigma\,\mathrm{d}\theta^2  +\frac{\sin^2\theta}{\Sigma}\left[(r^2+a^2)\mathrm{d}\phi -a\,\mathrm{d}t\right]^2 \,,
\end{equation}
where $\Delta$ and $\Sigma=\zeta\bar\zeta$ are defined in \eqref{eq:defs}
of the Letter.

The event horizon is a null hypersurface with $r=r_+$, with $r_+$ being the largest positive real root of $\Delta$. Furthermore, $M=\frac{r_+^2+a^2}{2r_+}$ is the ADM mass of the black hole, $J=M a$ is its ADM angular momentum, and $|a|\leq M$, with equality saturating at extremality, where the event horizon becomes degenerate. 
To present our results we also use $\chi=J/M^2$. The temperature and angular velocity of the Kerr solution are $T_H=\frac{r_+^2-a^2}{4\pi r_+(r_+^2+a^2)}$ and $\Omega_H=\frac{a}{r_+^2+a^2}$.

The line element of the 5-dimensional (5D) rotating Kerr string is $\mathrm{d}s^2_{\hbox{\tiny string}}=\mathrm{d}s^2_{\hbox{\tiny Kerr}}+\mathrm{d}z^2$.

For $\mu\neq0$, the four-dimensional massive system is equivalent to a vacuum perturbation of this string proportional to $e^{{\rm i}\mu z}$, with $H_{zA}=0$. We use this description when it makes the geometric origin of an identity more transparent; the resulting operators act on the four-dimensional field, and their symmetry properties can be checked directly in four dimensions.

%%%%%%%%%
\section{Instability growth rates and timescales}\label{sm:s:timescales}

The dominance diagram in Fig.~\ref{fig:growth} of the Letter identifies the fastest-growing branch among those compared in the plane of Kerr spin and graviton mass. Here we give the frequency surfaces underlying this comparison. We use $\alpha=\mu M$ and $\chi=J/M^2$, and write $\omega=\omega_R+{\rm i}\Gamma$. With our convention $e^{-{\rm i}\omega t}$, an unstable mode has $\Gamma>0$ and its amplitude grows as $e^{\Gamma t}$. Its e-folding time is therefore
\begin{equation*}
\frac{\tau_{\rm inst}}{M}=\frac{1}{M\Gamma}=\frac{1}{\operatorname{Im}(\omega M)}.
\end{equation*}
For an oscillatory mode the period is $T_{\rm osc}/M=2\pi/|\operatorname{Re}(\omega M)|$. The real and imaginary parts of the frequency thus determine different timescales.

\begin{figure}[!htbp]
\centering
\includegraphics[width=0.85\textwidth]{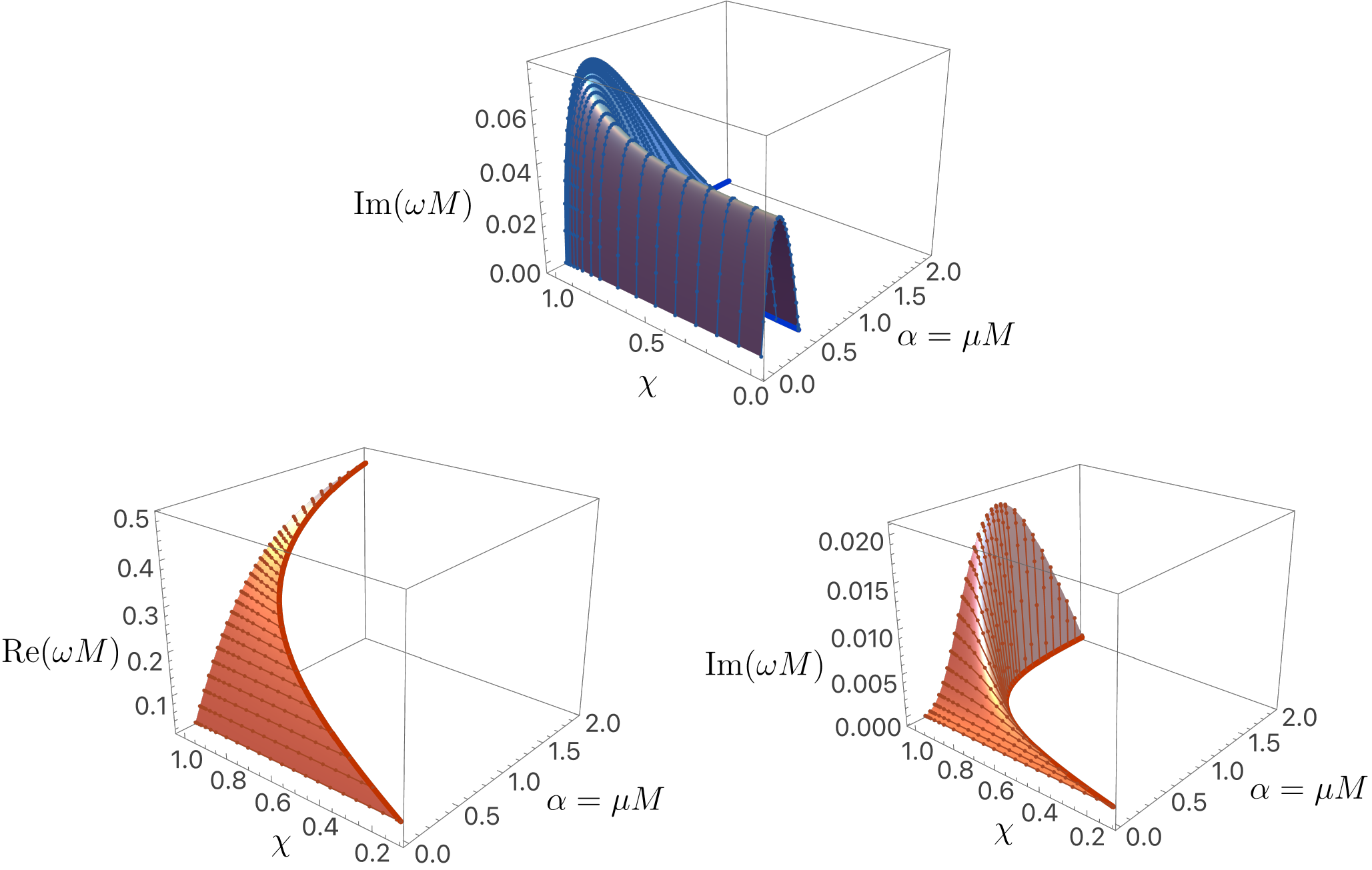}
\caption{Frequencies of the Gregory--Laflamme and dipole instabilities as functions of $\alpha=\mu M$ and $\chi=J/M^2$. Top panel: $\operatorname{Im}(\omega M)$ for the axisymmetric Gregory--Laflamme mode, whose real frequency vanishes. Bottom panel: $\operatorname{Re}(\omega M)$ (left plot) and $\Gamma\,M\equiv \operatorname{Im}(\omega M)$ (right plot) for the $\ell=m=1$ superradiant dipole branch. The inverse of the dimensionless growth rate gives $\tau_{\rm inst}/M$.}
\label{sm:fig:gl-dipole}
\end{figure}

Fig.~\ref{sm:fig:gl-dipole} shows the axisymmetric Gregory--Laflamme branch (top panel) and the superradiant dipole branch labelled by $\ell=m=1$ (bottom panel). The Gregory--Laflamme mode is purely growing, with $\omega_R=0$, and is the only unstable branch considered here that persists in the Schwarzschild limit. The lower panel shows both the oscillation frequency and the growth rate of the dipole mode. Its growth rate $\Gamma$ increases towards rapid rotation over much of the displayed unstable domain, and tends to zero at the onset, where $\omega_R=\Omega_H$. Correspondingly, the growth time diverges as the onset is approached, even when the oscillation period remains finite.

\begin{figure}[!htbp]
\centering
\includegraphics[width=0.85\textwidth]{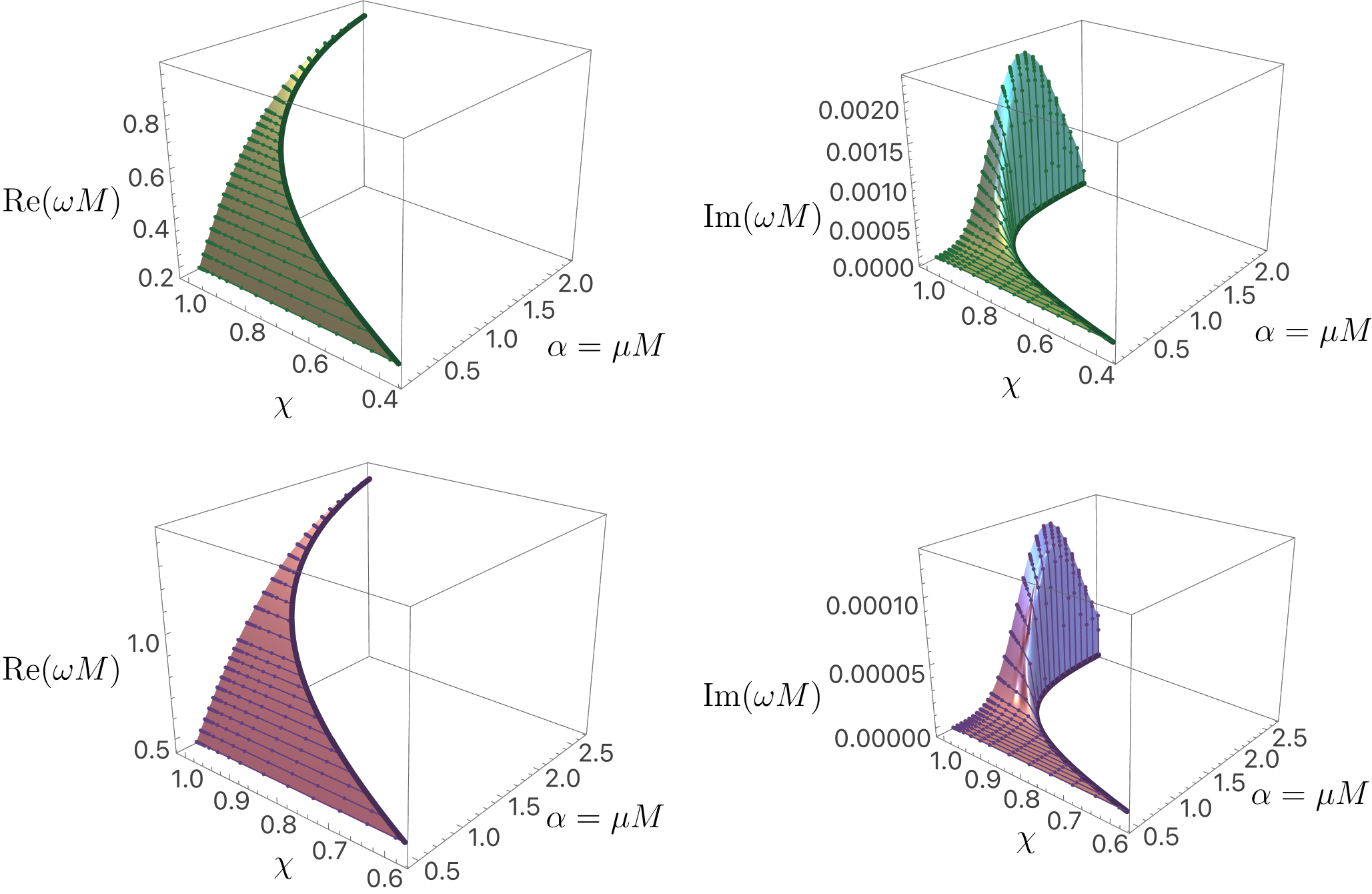}
\caption{Real (left plot) and imaginary (right plot) parts of $\omega M$ for the superradiant branches labelled by $\ell=m=2$ (top panel) and $\ell=m=3$ (bottom panel), as functions of $\alpha=\mu M$ and $\chi=J/M^2$. The different vertical scales in the right-hand plots display the hierarchy of growth rates. Positive $\operatorname{Im}(\omega M)$ gives an instability with $\tau_{\rm inst}/M=1/\operatorname{Im}(\omega M)$.}
\label{sm:fig:higher-m}
\end{figure}

Fig.~\ref{sm:fig:higher-m} gives the corresponding frequency surfaces for the $\ell=m=2$ (top panel) and $\ell=m=3$ (bottom panel) superradiant branches. The vertical scales of the imaginary parts show the progressively longer growth times of these branches near their respective largest plotted rates. For reference, $M\Gamma=10^{-2},10^{-3},10^{-4}$ corresponds to $\tau_{\rm inst}/M=10^2,10^3,10^4$, respectively. Nevertheless, a branch with a smaller peak growth rate can dominate where the faster branches are stable. The comparison must therefore be made at the same $(\chi,\alpha)$: the branch selected in Fig.~\ref{fig:growth} of the Letter has the shortest growth time among the four branches considered there. These are linear amplitude growth times; the time required to reach a given amplitude also depends on the initial perturbation.

The growth rates decrease sharply away from their maxima and vanish at the superradiant onset. Resolving these small imaginary parts was a limitation of our earlier PDE calculations, especially when comparing with small-spin expansions. The separated formulation makes this regime substantially more accessible numerically.

The Kerr branches considered here remain bound and superradiantly unstable towards $\alpha\to0$ at fixed nonzero spin, with both $\omega_R$ and $\Gamma$ tending to zero. There is therefore no additional marginal-binding edge at finite $\alpha$. Unstable modes can also occur at arbitrarily small rotation, provided the mass coupling is reduced accordingly.

This differs from the six-dimensional equal-angular-momenta Myers--Perry string~\cite{Dias:2022mde}, where the unstable region is bounded both by the superradiant onset, $\omega_R=m\Omega_H$, and by a marginal-binding edge, $\omega_R=\mu$, with $\Gamma=0$ on both boundaries. We interpret this contrast in terms of the different trapping properties: Kerr has a long-range $1/r$ gravitational potential and stable timelike orbits, whereas the five-dimensional Myers--Perry black hole underlying the six-dimensional string has no such stable orbits.

%\clearpage 
%%%%%%%%%%%%%%%%%%%%%%%%%%%%%%%%%%
\section{Coordinate operators and covariant symmetries}\label{sm:operators}
\subsection{Angular and radial operators}
Here we give the coordinate operators used in the Letter. They act on the five angular components in the order $s=2,1,0,-1,-2$.
Introducing the quantities
\begin{equation}\label{eq:LsQd}
Q=\frac{m}{\sin\theta}-a\omega\sin\theta, \qquad L_n=\partial_\theta+Q+n\cot\theta, \qquad L_n^\dagger=\partial_\theta-Q+n\cot\theta, \qquad  d_s=-{\rm i}(as\cos\theta+s^2\omega/\mu^2)\,,
\end{equation}
the first angular operator is a first-order spin ladder,
\begin{equation}\label{eq:K1}
K_1=\begin{pmatrix}
d_2 & L^\dagger_{-1} & 0 & 0 & 0\\
\mu^{-2}L_2 & d_1 & -\tfrac32L^\dagger_0 & 0 & 0\\
0 & -\mu^{-2}L_1 & 0 & -\mu^{-2}L^\dagger_1 & 0\\
0 & 0 & -\tfrac32 L_0 & d_{-1} & \mu^{-2}L^\dagger_2\\
0 & 0 & 0 & L_{-1} & d_{-2}
\end{pmatrix}.
\end{equation}
The second angular operator is a spin-weighted spheroidal operator with algebraic couplings.
Its diagonal elements are
\begin{align}\label{eq:K2}
(K_2)_{ss}&=\frac{\partial_\theta(\sin\theta\,\partial_\theta)}{\sin\theta}-\frac{(m+s\cos\theta)^2}{\sin^2\theta} +a^2(\omega^2-\mu^2)\cos^2\theta-2sa\omega\cos\theta-s^2 ,
\end{align}

and its off-diagonal entries are ${\rm i}a\sin\theta\,\kappa_{ss'}$, with the only nonzero constants $\kappa_{ss'}$ being
\begin{gather*}
\kappa_{21}=-2\mu^2,\ \ \kappa_{12}=2,\ \ \kappa_{10}=3\mu^2,\ \ \kappa_{01}=-2, \quad \hbox{and} \quad \kappa_{-s,-s'}=-\kappa_{s,s'}\,.
\end{gather*}

The radial operator is the spin-2 Teukolsky operator with a mass term,
\begin{equation}\label{eq:D}
\Dr=\Delta^{-2}\partial_r\big(\Delta^3\partial_r\big)+\frac{K^2-2{\rm i}K\Delta'}{\Delta}+8{\rm i}\omega r-\mu^2r^2\\
+6+2am\omega-a^2\omega^2 .
\end{equation}

\subsection{Covariant symmetry operators}
Let $f_{ab}$ be the Killing-Yano tensor of Kerr, $k_{ab}=({\star f})_{ab}$ the principal tensor, so that $\nabla_ck_{ab}=-2g_{c[a}\xi_{b]}$ and $K_{ab}=f_{ca}f^c{}_b$ is Carter's Killing tensor~\cite{Carter:1968rr,Frolov:2017kze}. Define the principal-tensor symmetry operator  
\begin{align} 
\mathcal K_1h_{ab}&=-2k_{(a}{}^ch_{b)c}+\frac{2}{\mu^2}\nabla_{(a}W_{b)}+\frac{4}{\mu^2}\mathcal L_\xi h_{ab},\label{eq:calK1}\\
W_b&=k^{cd}\nabla_ch_{bd}-2\xi^ch_{bc},\nonumber
\end{align}
and the spin-2 Carter operator
\begin{align}
\mathcal K_2 h_{ab} &=\nabla_c\big(K^{cd}\nabla_dh_{ab}\big)-2f_{(a}{}^{c}f^{de}[\nabla_c,\nabla_d]h_{b)e}
+4\xi^d\big(k_{(a}{}^{c}\nabla_{|c|}h_{b)d}-2k_{(a}{}^{c}\nabla_{|d|}h_{b)c}\big)\nonumber\\
&\quad+4\xi_{(a}k^{cd}\nabla_{|c|}h_{b)d}+4\xi^ck_{cd}\nabla_{(a}h_{b)}{}^{d}
-8\xi_{(a}\xi^ch_{b)c}+6\,\xi^2h_{ab}+2g_{ab}\,\xi^c\xi^dh_{cd}
-\mathcal L_\xi\big(a^2\mathcal L_\xi+2a\mathcal L_\eta\big)h_{ab},\label{eq:calK2}
\end{align}
with $\eta=\partial_\phi$. 
Both operators map Fierz--Pauli solutions to solutions. Their actions on the five components give Eq.~(8).
The identities proving these statements are given in Sec.~\ref{sm:s:sym}.

\subsection{Adjoint equation for the radial polynomial}
The polynomial $A(r)$ used in
Eq.~(13) solves
\begin{equation}\label{eq:Aeq}
\Delta A'''-(\Delta'+2{\rm i}K)A''+\big(\nu+2+2am\omega-a^2\omega^2\\
+6{\rm i}\omega r-\mu^2r^2\big)A'-\big(2\mu^2\lambda-4\mu^2r+8{\rm i}\omega\big)A=0 .
\end{equation}
Its coefficients and the termination condition are given in Sec.~\ref{sm:s:radial}.

%%%%%%%%%%%%%%%%%%%%%%%%%%%%%%%
\section{Conventions and tools}\label{sm:s:tools}

\paragraph{Background.} We use the Kinnersley tetrad on Kerr~\cite{Teukolsky:1972my,Teukolsky:1973ha}, signature $(-,+,+,+)$ with $l\cdot n=-1$, $m\cdot\bar m=1$, Wald's sign for the Riemann tensor, $\nabla_a\nabla_b\omega_c-\nabla_b\nabla_a\omega_c=R_{abc}{}^d\omega_d$, and GHP notation~\cite{Geroch:1973am}. In this tetrad $\kappa=\sigma=\epsilon=0$ and
\begin{equation}
\rho=-\frac1\zeta,\qquad \tau=-\frac{{\rm i}a\sin\theta}{\sqrt2\,\Sigma},\qquad \Psi_2=-\frac{M}{\zeta^3},\qquad \zeta=r-{\rm i}a\cos\theta,\quad \Sigma=\zeta\bar\zeta .
\end{equation}
The functions $\zeta,\bar\zeta$ have GHP type $(p,q)=(0,0)$, and hence zero boost and spin weights.
They obey
\begin{equation}\label{sm:eq:zeta}
\thorn\zeta=-\rho\zeta=1,\qquad \thorn'\zeta=-\rho'\zeta,\qquad \eth\zeta=-\tau\zeta,\qquad \eth'\zeta=-\tau'\zeta,
\end{equation}
together with the conjugate relations ($\thorn\bar\zeta=-\bar\rho\bar\zeta=1$, $\eth\bar\zeta=-\bar\tau'\bar\zeta$, $\eth'\bar\zeta=-\bar\tau\bar\zeta$, $\thorn'\bar\zeta=-\bar\rho'\bar\zeta$). When taking a further GHP derivative, the products $-\rho\zeta$ and $-\bar\rho\bar\zeta$ must retain their GHP types, even though both equal $1$ in the Kinnersley frame.

\paragraph{Spin coefficients.} The derivatives of the spin coefficients and Weyl scalars used throughout are
\begin{gather*}
\thorn\rho=\rho^2,\qquad \thorn\tau=\rho(\tau-\bar\tau'),\qquad \thorn\tau'=2\rho\tau',\qquad \thorn\Psi_2=3\rho\Psi_2,\qquad
\thorn\rho'=\thorn'\rho=(2\rho+\bar\rho)\rho'+\tfrac{\rho+\bar\rho}{2\rho\zeta^2}+\tfrac{\bar\rho}{\rho}\Psi_2,\\
\thorn'\tau=2\rho'\tau,\qquad \thorn'\tau'=\rho'(\tau'-\bar\tau),\qquad \thorn'\rho'=\rho'^2,\qquad \thorn'\Psi_2=3\rho'\Psi_2,\\
\eth\rho=(\rho-\bar\rho)\tau,\qquad \eth\tau=\tau^2,\qquad \eth\rho'=2\rho'\tau,\qquad \eth\Psi_2=3\tau\Psi_2,\qquad
\eth\tau'=\eth'\tau=\tau\tau'+(\rho-\bar\rho)\rho'+\tfrac12\Psi_2-\tfrac{\rho}{2\bar\rho}\bar\Psi_2,\\
\eth'\rho=2\rho\tau',\qquad \eth'\tau'=\tau'^2,\qquad \eth'\rho'=(\rho'-\bar\rho')\tau',\qquad \eth'\Psi_2=3\tau'\Psi_2,
\end{gather*}
and their complex conjugates.

\paragraph{Commutators.} On a quantity of GHP type $(p,q)$,
one has
\begin{align*}
[\thorn,\thorn']&=(\bar\tau-\tau')\eth+(\tau-\bar\tau')\eth'+p(\tau\tau'-\Psi_2)+q(\bar\tau\bar\tau'-\bar\Psi_2),\\
[\eth,\eth']&=(\bar\rho'-\rho')\thorn+(\rho-\bar\rho)\thorn'+p(\rho\rho'+\Psi_2)-q(\bar\rho\bar\rho'+\bar\Psi_2),\\
[\thorn,\eth]&=\bar\rho\,\eth-\bar\tau'\thorn+q\,\bar\rho\bar\tau',\qquad [\thorn',\eth]=\rho'\eth-\tau\thorn'-p\,\rho'\tau,
\end{align*}
and the conjugates $[\thorn,\eth']=\rho\,\eth'-\tau'\thorn+p\,\rho\tau'$, $[\thorn',\eth']=\bar\rho'\eth'-\bar\tau\thorn'-q\,\bar\rho'\bar\tau$.

\paragraph{Killing vectors and modes.} The two Killing vectors of Kerr are $\xi=\partial_t$ and $\eta=\partial_\phi$. The tetrad components of $\xi$ have GHP types $(1,1)$, $(-1,-1)$, $(1,-1)$ and $(-1,1)$, respectively. We write them in a form that retains these types when taking further GHP derivatives:
\begin{equation}
\xi_l=\zeta\rho\ (=-1),\qquad \xi_n=-\bar\zeta\bar\rho',\qquad \xi_m=-\bar\zeta\bar\tau',\qquad \xi_{\bar m}=-\zeta\tau',
\end{equation}
and the Lie derivative along $\xi$ of a GHP quantity of type $(p,q)$ is
\begin{equation}\label{sm:eq:Lxi}
\mathcal L_\xi=\bar\zeta\bar\rho'\thorn-\zeta\rho\,\thorn'-\zeta\tau'\eth-\bar\zeta\bar\tau'\eth'-\tfrac12\big(p\,\zeta\Psi_2+q\,\bar\zeta\bar\Psi_2\big).
\end{equation}
The products multiplying the four GHP derivatives in \eqref{sm:eq:Lxi}, together with its $(p,q)$-dependent algebraic term, give the Lie derivative on a weighted quantity. With the type shifted appropriately after each GHP derivative, $\mathcal L_\xi$ commutes with $\thorn,\thorn',\eth,\eth'$. On Fourier modes $e^{-{\rm i}\omega t+{\rm i}m\phi}$, $\mathcal L_\xi=-{\rm i}\omega$ and $\mathcal L_\eta={\rm i}m$.

\section{The transport identity}\label{sm:s:transport}

\subsection{Field equations and the curl hierarchy}
With the transverse and traceless constraints, $h\equiv h^a{}_a=0$ and $Y_a\equiv\nabla^bh_{ab}=0$, the Fierz--Pauli system (1) %\eqref{eq:FP}
can be written as
\begin{equation}\label{eq:FP-Leom}
\mathfrak L_{ab}\equiv-\Box h_{ab}-2R_{acbd}h^{cd}+\mu^2h_{ab}=0,\qquad Y_a=0,\qquad h=0 ,
\end{equation}
where the first two terms form the Lichnerowicz operator.
Off shell we use the four-dimensional field-equation and constraint data $(E,Z,h)$:
\begin{equation}
E_{ab}=\delta R_{ab}+\tfrac12\mu^2h_{ab}=\tfrac12\mathfrak L_{ab}+\nabla_{(a}v_{b)},\qquad Z_a=\nabla^bh_{ab}-\nabla_ah,\qquad h,
\end{equation}
with $v=Y-\tfrac12\nabla h$. For $\mu\neq0$, $E_{ab}=0$, $Z_a=0$ and $h=0$ are equivalent to the Fierz--Pauli system. For comparison with the string interpretation, these quantities give the Fourier amplitudes $\delta\widehat R_{ab}=E_{ab}$, $\delta\widehat R_{az}=\tfrac{{\rm i}\mu}{2}Z_a$ and $\delta\widehat R_{zz}=\tfrac{\mu^2}{2}h$ of $H=h\,e^{{\rm i}\mu z}$ with $H_{zA}=0$.

The curl defined in (4) %\eqref{eq:field-components}
$F_{abc}\equiv \nabla_ah_{bc}-\nabla_bh_{ac}$,
obeys three identities that follow from the Ricci identity and the definition of $\delta R_{ab}$:
\begin{align}
&\nabla_aF_{bcd}+\nabla_bF_{cad}+\nabla_cF_{abd}=\textstyle\sum_{\rm cyc(abc)}R_{abd}{}^eh_{ce}\qquad\text{(cyclic)},\label{sm:eq:cyc}\\
&\nabla^aF_{abd}=-2\,\delta R_{bd}+\nabla_dZ_b+R_{adbe}h^{ae}\qquad\text{(divergence)},\label{sm:eq:div}\\
&\nabla_a(\xi\cdot h)_b-\nabla_b(\xi\cdot h)_a=\xi^cF_{abc}+(\nabla_a\xi^c)h_{bc}-(\nabla_b\xi^c)h_{ac}\qquad\text{(Killing)}.\label{sm:eq:kil}
\end{align}
A further derivative of $F$, together with the background-curvature terms from commuting covariant derivatives, gives the linearised Riemann tensor. The components of $\vO$ in (3) %\eqref{eq:Omega}
are one from each layer: $h_{ll}$, $X_1=F_{lml}$ and $\Psi_0=\delta C_{lmlm}$ and the $m\leftrightarrow\bar m$ partners of the latter two. In GHP form,
\begin{gather}
X_1=(\thorn-2\bar\rho)h_{lm}-(\eth-\bar\tau')h_{ll},\\
\Psi_0=-\tfrac12\Big[(\eth-\bar\tau')^2h_{ll}+(\thorn-\bar\rho)^2h_{mm}-(\thorn-\bar\rho)(\eth-2\bar\tau')h_{lm}-(\eth-\bar\tau')(\thorn-2\bar\rho)h_{lm}\Big],
\end{gather}
and the other curl components that appear below are
\begin{align*}
F_{lmm}&=(\thorn-\bar\rho)h_{mm}-(\eth-2\bar\tau')h_{lm},\\
F_{lm\bar m}&=(\thorn-\bar\rho)h_{m\bar m}-\bar\rho\,h_{ln}-(\eth-\bar\tau')h_{l\bar m}+\tau'h_{lm}-\rho'h_{ll},\\
F_{l\bar mm}&=(\thorn-\rho)h_{m\bar m}-\rho\,h_{ln}-(\eth'-\tau')h_{lm}+\bar\tau'h_{l\bar m}-\bar\rho'h_{ll},\\
F_{lmn}&=(\thorn-\bar\rho)h_{nm}-(\eth-\bar\tau')h_{ln}+\bar\tau'h_{m\bar m}+\tau'h_{mm}-\rho'h_{lm},\\
F_{lnm}&=\thorn h_{nm}-\thorn'h_{lm}+(\bar\tau'-\tau)(h_{ln}+h_{m\bar m})+(\tau'-\bar\tau)h_{mm},\\
F_{m\bar mm}&=\eth h_{m\bar m}-\eth'h_{mm}+(\rho'-2\bar\rho')h_{lm}+(\bar\rho-2\rho)h_{nm}.
\end{align*}

\subsection{Statement of the transport identity}\label{sm:s:transport-Statement}
Write $c_s=(2,\sqrt2,1,\sqrt2,2)$ for $s=2,\dots,-2$, so that $\vO_s=c_s\times(\Psi_0,X_1,h_{ll},\bar X_1,\bar\Psi_0)$, and let $\mathcal E$ be the transport operator (9) %\eqref{eq:transport}
of the Letter.

\emph{Theorem.} For every symmetric $h$, and each $s$,
\begin{equation}\label{sm:eq:thm}
\frac{\mathcal E_s[\vO[h]]}{c_s}=\thorn\big(\Sigma\,Y_s[R]\big)+N_s[R],
\end{equation}
where $R=(E,Z,h)$ denotes the four-dimensional off-shell data defined above, with $R_{ll}=E_{ll}$, $R_{lm}=E_{lm}$ and $R_{mm}=E_{mm}$:
\begin{align}\label{sm:eq:thmAUX}
Y_2&=-\mathcal S_0,\qquad N_2=0,\nonumber\\
\mathcal S_0&=(\eth-\bar\tau'-4\tau)\big[(\thorn-2\bar\rho)R_{lm}-(\eth-\bar\tau')R_{ll}\big]+(\thorn-4\rho-\bar\rho)\big[(\eth-2\bar\tau')R_{lm}-(\thorn-\bar\rho)R_{mm}\big],\nonumber\\
Y_1&=2(\eth-\bar\tau'-3\tau)R_{ll}-2(\thorn-3\rho-3\bar\rho)R_{lm}+(\thorn-\rho-\bar\rho)\big[(\thorn-2\rho-\bar\rho)Z_m-(\eth-2\tau-\bar\tau')Z_l\big],\nonumber\\
N_1&=-\tfrac{2}{\rho}\big[(\eth-\bar\tau'-3\tau)R_{ll}+2(\rho+\bar\rho)R_{lm}\big],\nonumber\\
Y_0&=-2R_{ll}+2(\thorn-2\rho-2\bar\rho)Z_l+(\thorn-\rho-\bar\rho)^2h,\qquad N_0=4\big(\rho^{-1}+\bar\rho^{-1}\big)R_{ll}=-8r\,R_{ll}.
\end{align}
$\mathcal S_0$ is Teukolsky's source operator. The rows $s=-1,-2$ follow from the mirror $m\leftrightarrow\bar m$ ($\eth\leftrightarrow\eth'$, $\rho\leftrightarrow\bar\rho$, $\tau\leftrightarrow\bar\tau$, $\tau'\leftrightarrow\bar\tau'$, $R_{lm}\leftrightarrow R_{l\bar m}$, \dots), under which all the tools of Sec.~\ref{sm:s:tools} are invariant. On a Fierz--Pauli solution $R=0$, so the right-hand side of \eqref{sm:eq:thm} vanishes, which gives the closed transport equation (9) %\eqref{eq:transport} 
of the Letter.

On the string, with $\eth_z\equiv\partial_z={\rm i}\mu$, one has $Y_2=-\mathcal S_{mm}$, $Y_1=-\tfrac2{\eth_z}\mathcal S_{mz}$ and $Y_0=\tfrac2{\eth_z^2}\mathcal S_{zz}$. Here
\begin{align*}
\mathcal S_{mz}&=\eth_z\big[(\thorn-3\rho-3\bar\rho)R_{lm}-(\eth-\bar\tau'-3\tau)R_{ll}\big]+(\thorn-\rho-\bar\rho)\big[(\eth-2\tau-\bar\tau')R_{lz}-(\thorn-2\rho-\bar\rho)R_{mz}\big],\\
\mathcal S_{zz}&=\eth_z\big[(\thorn-3\rho-3\bar\rho)R_{lz}-\eth_zR_{ll}\big]+(\thorn-\rho-\bar\rho)\big[\eth_zR_{lz}-(\thorn-\rho-\bar\rho)R_{zz}\big]
\end{align*}
are the $(mz)$ and $(zz)$ components of a Teukolsky-type source operator on the string. The flat $z$ direction carries no spin coefficients, and the expansion $\rho+\bar\rho$ replaces $4\rho+\bar\rho$.

\subsection{Structure of the proof}

For each row, the proof of \eqref{sm:eq:thm} combines two identities, both valid for every symmetric $h$:
\begin{align}
(T_s):\qquad &\frac{[(\Dr+K_2)\vO]_s}{c_s}=\Sigma\,\mathcal C_s[\mathfrak L]+\Sigma\,P_s[Y,h]+G_s, \label{eq:Ts}\\
(B_s):\qquad &\thorn G_s=N_s[R]-\frac{\big[(2\mu^2K_1-4\mu^2r+8{\rm i}\omega)\vO\big]_s}{c_s}.\label{eq:Bs}
\end{align}

The first identity follows from the field equations and constraints before setting them to zero. For $(T_0)$ one uses the $ll$ equation, $Y_l$ and $h$; for $(T_1)$ one takes differentiated combinations of the $ll$ and $lm$ equations and the constraints. For $(T_2)$ one uses the off-shell Teukolsky identity, whose source consists of the $ll,lm,mm$ components of the linearised Einstein tensor, and substitutes \eqref{eq:FP-Leom}. The detailed derivations are given below. Here $\mathcal C_s$ is a differential projection of $\mathfrak L$, $P_s$ is linear in $Y$ and $h$, and $G_s$ obstructs closure at second order because it contains components of $F$ and $\xi\cdot h$ outside $\vO$. The second identity gives a first-order equation for $G_s$. Differentiating $(T_s)$ and using $(B_s)$ eliminates $G_s$ and gives \eqref{sm:eq:thm}. The projections are
\begin{equation}\label{eq:Cs}
\mathcal C_0[\mathfrak L]=-\mathfrak L_{ll},\qquad \mathcal C_1[\mathfrak L]=(\eth-\bar\tau'-3\tau)\mathfrak L_{ll}-(\thorn-3\rho-3\bar\rho)\mathfrak L_{lm},\qquad \mathcal C_2[\mathfrak L]=-\tfrac12\mathcal S_0[\mathfrak L].
\end{equation}
and
\begin{align}\label{eq:Ps}
P_0 &=-4(\rho+\bar\rho)Y_l+2(\rho+\bar\rho)\thorn h+2\rho\bar\rho\,h \,,\nonumber \\
P_1 &=\tfrac{2\tau}{\bar\zeta}Y_l+\tfrac{2(\zeta+3\bar\zeta)}{\zeta\bar\zeta^2}Y_m-4\tau\thorn Y_l-\tfrac{2(\zeta+2\bar\zeta)}{\zeta\bar\zeta}\eth Y_l-\tfrac{3\tau-\bar\tau'}{\bar\zeta}\thorn h-\tfrac{2}{\Sigma}\eth h+\tau\thorn^2h+\tfrac{\zeta+\bar\zeta}{\Sigma}\thorn\eth h\,,\nonumber \\
P_2 &=-\mathcal S_0[\nabla_{(a}v_{b)}],\qquad v_a=Y_a-\tfrac12\nabla_a h.
\end{align}

The following \emph{separation lemma} converts the GHP wave operators obtained from the component equations into radial and angular operators. It is used in each derivation of $(T_s)$ below; the off-diagonal angular terms are then supplied by the couplings between components of $\vO$. Let $(p,q)=(2+s,2-s)$. On any quantity of this type,
\begin{equation}\label{sm:eq:seplemma}
\Sigma^{-1}(\Dr+K_{2,ss})+\mu^2=-2\thorn\thorn'+2\eth\eth'+2\rho'\thorn-2(\tau'+q\bar\tau)\eth+2\big(p\rho+(q+1)\bar\rho\big)\thorn'-2(p\tau+\bar\tau')\eth'+U_s,
\end{equation}
with
\begin{equation*}
U_2=4\Psi_2-8\rho\rho'+8\tau\tau',\qquad U_1=-6(\rho+\bar\rho)\rho'+6\tau\tau'+2\bar\tau\bar\tau',\qquad U_0=-2\Psi_2+2\bar\Psi_2-4\rho\rho'-8\bar\rho\rho'+4\tau\tau'+4\bar\tau\bar\tau' .
\end{equation*}
For $s=2$, the entire right-hand side of \eqref{sm:eq:seplemma} is $-2\mathcal T$, where $\mathcal T=(\thorn-4\rho-\bar\rho)(\thorn'-\rho')-(\eth-4\tau-\bar\tau')(\eth'-\tau')-3\Psi_2$ is the spin-2 Teukolsky operator in the normalisation used below. Thus $\Dr+K_{2,22}=-2\Sigma\mathcal T-\mu^2\Sigma$. For $s=0,1$, the same lemma identifies the corresponding boost-weight-two wave operators, with the potentials $U_0,U_1$ written above.
The lemma \eqref{sm:eq:seplemma} is checked by inserting the Kinnersley tetrad on modes. It says that the boost-weight-two, spin-$s$ wave operator separates into $\Dr+K_{2,ss}$. The off-diagonal entries of $K_2$ are ${\rm i}a\sin\theta=-\sqrt2\,\Sigma\tau$ times the constants $\kappa_{ss'}$ in Sec.~\ref{sm:operators}.

The next three subsections derive \eqref{eq:Ts} and \eqref{eq:Bs} for the rows $s=2,0,1$. The rows $s=-2,-1$ follow by the mirror operation specified above, with the frequency and azimuthal number held fixed.

\subsection{The row \texorpdfstring{$s=2$}{s=2}}

For $(T_2)$ we start from Teukolsky's identity in Wald's off-shell form, $\mathcal T\Psi_0=\tfrac12\mathcal S_0[\delta G]$, with $\mathcal T$ and $\mathcal S_0$ defined above. The source operator depends only on $\delta G_{ll},\delta G_{lm},\delta G_{mm}$; these equal the corresponding Ricci components because the trace term has zero projection on these tetrad pairs. We then insert $\delta R_{ab}=\mathfrak L_{ab}/2+\nabla_{(a}v_{b)}-\mu^2h_{ab}/2$. The first two terms give $\mathcal C_2$ and $P_2$.

Expanding the source operator gives $\mathcal S_0[h]=2\Psi_0-4\tau X_1+4\rho F_{lmm}$. The mass contribution $-\tfrac14\mu^2\mathcal S_0[h]$ therefore splits into three terms.
\begin{itemize}
\item $-\tfrac12\mu^2\Psi_0$, the diagonal mass term;
\item $\mu^2\tau X_1$, the coupling to $X_1$ in $K_2$;
\item the remaining curl term $-\mu^2\rho F_{lmm}$. Multiplying the Teukolsky identity by $-2\Sigma$ converts it to the normalisation of $(T_2)$ and gives $G_2=2\mu^2\Sigma\rho F_{lmm}=-2\mu^2\bar\zeta F_{lmm}$.
\end{itemize}
The $lmm$ component of the cyclic identity \eqref{sm:eq:cyc} reads
\begin{equation}\label{sm:eq:B2}
(\thorn-\bar\rho)F_{lmm}=(\eth-\bar\tau')X_1-2\Psi_0 .
\end{equation}

Multiplying by $\bar\zeta$ and using $\thorn\bar\zeta=1$ and $\bar\zeta\bar\rho=-1$ from \eqref{sm:eq:zeta}, it follows that $\thorn(\bar\zeta F_{lmm})=\bar\zeta\big[(\eth-\bar\tau')X_1-2\Psi_0\big]$. Hence $\thorn G_2=-2\mu^2\bar\zeta\big[(\eth-\bar\tau')X_1-2\Psi_0\big]$. 
Using $d_2-2r+4{\rm i}\omega/\mu^2=-2\bar\zeta$ from \eqref{eq:LsQd}, we recover $(B_2)$ in \eqref{eq:Bs}, with $N_2=0$.
Finally, applying $\thorn$ to $(T_2)$ and using $(B_2)$ gives \eqref{sm:eq:thm} with the extra curl component eliminated.

%%%%%
\subsection{The row \texorpdfstring{$s=0$}{s=0}}

To derive $(T_0)$, we evaluate $\mathfrak L_{ll}$, $Y_l$ and $h$ in \eqref{eq:FP-Leom}, using the following component formulas in the Kinnersley tetrad:

\begin{align*}
(\Box h)_{ll}&=\big[-2\thorn\thorn'+2\eth\eth'+2\rho'\thorn+2\bar\rho\thorn'-2\tau'\eth-2\bar\tau'\eth'+4\bar\rho\bar\rho'-4\Psi_2\big]h_{ll}\\
& \quad +4\rho\bar\rho\,(h_{ln}+h_{m\bar m})+4(\bar\rho\eth'-\bar\tau\thorn)h_{lm}+4(\rho\eth-\tau\thorn)h_{l\bar m},\\
R_{lcld}h^{cd}&=(\Psi_2+\bar\Psi_2)\,h_{ll},\\
Y_l&=-(\thorn'-\rho'-\bar\rho')h_{ll}-(\thorn-\rho-\bar\rho)h_{ln}+(\eth'-\tau'-2\bar\tau)h_{lm}+(\eth-\bar\tau'-2\tau)h_{l\bar m}+(\rho+\bar\rho)h_{m\bar m},
\end{align*}
together with $h=-2h_{ln}+2h_{m\bar m}$.

Together with $\Sigma\mathcal C_0[\mathfrak L]=-\Sigma\mathfrak L_{ll}$ from \eqref{eq:Cs}, these expressions give $(T_0)$ with 
$P_0$ given in \eqref{eq:Ps} and $G_0$ defined below.
The terms in $(T_0)$ can be grouped as follows:
\begin{itemize}
\item The second-order $h_{ll}$ terms form the boost-weight-two, spin-0 wave operator, which by \eqref{sm:eq:seplemma} is $\Dr+K_{2,00}$.
\item The $\eth'h_{lm}$ and $\eth h_{l\bar m}$ terms combine with the $\eth h_{ll}$ pieces of $X_1$ and $\bar X_1$ into the $K_2$ couplings $\propto {\rm i}a\sin\theta\,(X_1,\bar X_1)$.
\item What remains is the obstruction
\begin{equation}
G_0=4\big(\zeta F_{lm\bar m}+\bar\zeta F_{l\bar mm}\big)+8\,\xi^ch_{lc},
\end{equation}
whose algebraic part is $8\xi^ch_{lc}=8\big(h_{ln}+\bar\rho'\bar\zeta\,h_{ll}-\tau'\zeta\,h_{lm}-\bar\tau'\bar\zeta\,h_{l\bar m}\big)$.
\end{itemize}

To close the system, we derive $(B_0)$ by taking $\thorn G_0$, as in the $s=2$ case. For that, we start by multiplying  $\thorn F_{lm\bar m}$ and $\thorn F_{l\bar mm}$ $-$ as given below in \eqref{eq:pF} $-$ by $\zeta$ and $\bar{\zeta}$, respectively. Then we use   $\thorn\zeta=\thorn\bar\zeta=1$ in \eqref{sm:eq:zeta} and the divergence identity \eqref{sm:eq:div} for $\delta R_{ll}$ to obtain $\thorn(\zeta F_{lm\bar m})$ and $\thorn(\bar\zeta F_{l\bar mm})$. This yields $(B_0)$, i.e. \eqref{eq:Bs},  which is $\mu$-independent:
\begin{equation}
\thorn G_0=-8r\,\delta R_{ll}+2\big(L_1\vO_1+L_1^\dagger\vO_{-1}\big)-8{\rm i}\omega\,h_{ll}.
\end{equation}
Using $\delta R_{ll}=E_{ll}-\mu^2h_{ll}/2$, this is $(B_0)$ with $N_0=-8rE_{ll}$. On shell $E_{ll}=0$, so $N_0=0$ and $\delta R_{ll}=-\mu^2h_{ll}/2$.
Finally, applying $\thorn$ to $(T_0)$ and using $(B_0)$ eliminates $G_0$ and gives \eqref{sm:eq:thm} for $s=0$.

%%%%%%
\subsection{The row \texorpdfstring{$s=1$}{s=1}}

For $(T_1)$, we use differentiated combinations of $\mathfrak L_{ll}$ and $\mathfrak L_{lm}$, together with $Y_l,Y_m$ and $h$, as specified by $\mathcal C_1$ and $P_1$ above.

To evaluate these, this time we need the following three component formulas in the Kinnersley tetrad:
\begin{align*}
(\Box h)_{lm}&=\big[-2\thorn\thorn'+2\eth\eth'+2\rho'\thorn+2\bar\rho\thorn'-2\tau'\eth-2\bar\tau'\eth'+4\bar\rho\bar\rho'-4\Psi_2-4\bar\tau\bar\tau'\big]h_{lm}-4\tau\bar\tau'h_{l\bar m}+4\rho\bar\rho\,h_{nm}\\
&\quad+2\bar\rho'\eth h_{ll}-2\bar\tau'\thorn'h_{ll}-2(\tau\thorn-\rho\eth)(h_{ln}+h_{m\bar m})-2\bar\tau\thorn h_{mm}+2\bar\rho\eth'h_{mm},\\
R_{lcmd}h^{cd}&=(\Psi_2-2\bar\Psi_2)h_{lm},\\
Y_m&=-\thorn'h_{lm}-\thorn h_{nm}+\eth h_{m\bar m}+\eth'h_{mm}+(\rho'+2\bar\rho')h_{lm}-(\tau+\bar\tau')(h_{ln}+h_{m\bar m})-(\bar\tau+\tau')h_{mm}+(2\rho+\bar\rho)h_{nm}
\end{align*}

To proceed, one commutes the outer $\thorn,\eth$ inwards so that the third-order terms collect into $(\thorn-2\bar\rho)(\cdot)h_{lm}-(\eth-\bar\tau')(\cdot)h_{ll}$. By \eqref{sm:eq:seplemma} with $s=1$, these give $\Sigma^{-1}(\Dr+K_{2,11}+\mu^2\Sigma)X_1$. The terms $\eth^2h_{ll}$, $\thorn\eth h_{lm}$, $\thorn^2h_{mm}$ assemble into $\Psi_0$ and reproduce the $K_2$ coupling to $\Psi_0$, and the $\mu^2h_{ll}$ term gives the coupling to $h_{ll}$. 
This gives $(T_1)$ with $\mathcal C_1[\mathfrak L]$ defined in \eqref{eq:Cs} and $P_1$ in \eqref{eq:Ps}, 

together with the obstruction
\begin{align*}
G_1&=\mu^2(\zeta+3\bar\zeta)h_{lm}+8\mathcal L_\xi h_{lm}+4\zeta\eth'F_{lmm}-2\zeta\eth\big(F_{l\bar mm}+F_{lm\bar m}\big)+2\zeta\thorn F_{m\bar mm}-\tfrac{2(\zeta+\bar\zeta)}{\bar\zeta}F_{lmn}\\
&\quad+\tfrac{2(\zeta+2\bar\zeta)}{\bar\zeta}\big(F_{lnm}+F_{m\bar mm}\big)+\bar\tau'\big[4(\zeta-2\bar\zeta)F_{l\bar mm}+2(\zeta-\bar\zeta)F_{lm\bar m}\big]+\tfrac{4\tau'\zeta(\zeta-3\bar\zeta)}{\bar\zeta}F_{lmm}\\
&\quad+\tfrac{4(\zeta+\bar\zeta)}{\bar\zeta^2}(\xi\cdot h)_m+\tfrac{4\bar\tau'(\zeta-\bar\zeta)}{\bar\zeta}(\xi\cdot h)_l-\tfrac{4(\zeta+\bar\zeta)}{\bar\zeta}\eth(\xi\cdot h)_l+\tfrac{4(\zeta-\bar\zeta)}{\bar\zeta}\thorn(\xi\cdot h)_m,
\end{align*}
with $(\xi\cdot h)_l=-\zeta\rho\,h_{ln}+\bar\zeta\bar\rho'h_{ll}-\zeta\tau'h_{lm}-\bar\zeta\bar\tau'h_{l\bar m}$ and $(\xi\cdot h)_m=-\zeta\rho\,h_{nm}+\bar\zeta\bar\rho'h_{lm}-\zeta\tau'h_{mm}-\bar\zeta\bar\tau'h_{m\bar m}$.

To get $(B_1)$ in \eqref{eq:Bs}, we apply $\thorn$ to $G_1$. The cyclic and divergence identities give the $\thorn$-derivative of every curl component that occurs:
\begin{align}\label{eq:pF}
(\thorn-\bar\rho)F_{lmm}&=(\eth-\bar\tau')X_1-2\Psi_0,\nonumber \\
\thorn F_{lm\bar m}&=(\eth'-\tau')X_1+2\rho F_{lm\bar m}-\rho F_{l\bar mm}+\rho F_{lnl}-\delta R_{ll}-\Psi_2h_{ll},\nonumber \\
\thorn F_{l\bar mm}&=(\eth-\bar\tau')\bar X_1+2\bar\rho F_{l\bar mm}-\bar\rho F_{lm\bar m}+\bar\rho F_{lnl}-\delta R_{ll}-\bar\Psi_2h_{ll},\nonumber \\
\thorn F_{m\bar mm}&=\eth F_{l\bar mm}-\eth'F_{lmm}-2\bar\tau'F_{l\bar mm}+\bar\tau'F_{lm\bar m}+\tau'F_{lmm}-\bar\rho'X_1-\rho F_{lmn}\nonumber \\
&\quad+(\bar\rho-\rho)F_{lnm}+(\rho+\bar\rho)F_{m\bar mm}+(\Psi_2+2\bar\Psi_2)h_{lm},\nonumber \\
\thorn F_{lmn}&=\thorn'X_1+\eth'F_{lmm}-\eth F_{lm\bar m}-\tau F_{l\bar mm}+(2\tau-\bar\tau')F_{lm\bar m}+(\bar\rho'-\rho')X_1+(\bar\tau-\tau')F_{lmm}\nonumber \\
&\quad+(\rho-\bar\rho)F_{lmn}+\tau F_{lnl}+\rho F_{lnm}-\rho F_{m\bar mm}-2\,\delta R_{lm}-2\Psi_2h_{lm},\nonumber \\
\thorn F_{lnm}&=\eth F_{lnl}-\eth F_{lm\bar m}+\eth'F_{lmm}-\bar\tau'F_{l\bar mm}+\bar\rho'X_1-\tau'F_{lmm}+(\rho-2\bar\rho)F_{lmn}-\bar\tau'F_{lnl}\nonumber \\
&\quad+(\rho+2\bar\rho)F_{lnm}-\rho F_{m\bar mm}-2\,\delta R_{lm}+(2\bar\Psi_2-\Psi_2)h_{lm}.
\end{align}
The Killing identity \eqref{sm:eq:kil} handles $\thorn(\xi\cdot h)$, and $\mathcal L_\xi$ commutes with $\thorn$. All curl components other than those in $\vO$ cancel, and one is left with
\begin{equation}\label{sm:eq:B1}
\thorn G_1=N_1-4\zeta(\eth'-4\tau')\Psi_0+\mu^2(\zeta+3\bar\zeta)X_1+6\,\mathcal L_\xi X_1+3\mu^2\bar\zeta(\eth-2\bar\tau')h_{ll},
\end{equation}
with $N_1$ given in \eqref{sm:eq:thmAUX}.
The $\mu^2$ sector of \eqref{sm:eq:B1} can be simplified by substituting $X_1=(\thorn-2\bar\rho)h_{lm}-(\eth-\bar\tau')h_{ll}$ and $\thorn(\zeta+3\bar\zeta)=4$.
In the end, in the notation of Sec.~\ref{sm:s:sym}, the right-hand side is then $N_1$ minus the $s=1$ row of $(2\mu^2K_1-4\mu^2r+8{\rm i}\omega)\vO/\sqrt2$; thus \eqref{sm:eq:B1} is $(B_1)$ in \eqref{eq:Bs}. 

%%%%%%%%%%%%%
\section{Hidden symmetries and the angular operators}\label{sm:s:sym}

The angular operators $K_1,K_2$ are induced by covariant symmetries of the Fierz--Pauli equations. The principal closed conformal Killing-Yano tensor gives $\mathcal K_1$, while a completion of Carter's operator gives $\mathcal K_2$. We verify their symmetry properties and derive the intertwining relation~\eqref{eq:intertwine}, which identifies $K_1,K_2$ as their actions on $\vO$.

\subsection{Angular operators in GHP form}
On a quantity of type $(p,q)$, with $s=\tfrac12(p-q)$, the entries of $K_1$ and $K_2$ in Sec.~\ref{sm:operators} are
\begin{gather*}
L_s=\sqrt2\,\zeta(\eth'-p\tau'),\qquad L^\dagger_{-s}=\sqrt2\,\bar\zeta(\eth-q\bar\tau'),\qquad d_s=\tfrac{s}{2}(\zeta-\bar\zeta)+\tfrac{s^2}{\mu^2}\mathcal L_\xi,\\
K_{2,ss}=2\bar\zeta\big(\eth-(q+1)\bar\tau'\big)\zeta\big(\eth'-p\tau'\big)-\mathcal L_{\rm sh}+\tfrac{\mu^2}{4}(\zeta-\bar\zeta)^2+(2s-1)(\zeta-\bar\zeta)\mathcal L_\xi-s(s-1),
\end{gather*}
with $\mathcal L_{\rm sh}=\mathcal L_\xi(a^2\mathcal L_\xi+2a\mathcal L_\eta)$. On Fourier modes these reduce to the coordinate expressions in Sec.~\ref{sm:operators}. $L_s$ lowers and $L^\dagger_{-s}$ raises the spin weight, and both act on the spin-$s$ entry.

\subsection{Background tensors}
The Killing-Yano tensor $f$, the principal tensor $k={\star f}$ and Carter's tensor $K_{ab}=f_{ca}f^c{}_b$ have the nonzero tetrad components
\begin{equation}
f_{ln}=a\cos\theta,\quad f_{m\bar m}=-{\rm i}r\,; \qquad k_{ln}=r,\quad k_{m\bar m}={\rm i}a\cos\theta\,;\qquad K_{ln}=a^2\cos^2\theta,\quad K_{m\bar m}=r^2 .
\end{equation}
They obey $\nabla_{(a}f_{b)c}=0$ and
\begin{equation}\label{sm:eq:I}
\nabla_ck_{ab}=-(g_{ca}\xi_b-g_{cb}\xi_a),\qquad \nabla^cK_{cd}=-2\chi_d,\qquad K^{cd}W_{cade}=-(1+\xi^2)g_{ae}-4(Fk)_{ae},
\end{equation}
where $\chi_a=k_{ab}\xi^b$, $F_{ab}=\nabla_a\xi_b$ (not to be confused with the curl $F_{abc}$) and $(Fk)_{ae}=F_a{}^ck_{ce}$. In this section $W_{abcd}$ denotes curvature with the convention $[\nabla_a,\nabla_b]\omega_c=-W_{abc}{}^d\omega_d$. Thus, on the Ricci-flat background, $W_{abcd}=-C_{abcd}$ relative to the convention of Sec.~\ref{sm:s:tools}, with the same index ordering. Comparisons with the earlier formulas therefore require this sign translation.  The last identity in \eqref{sm:eq:I} is a Petrov type-D statement; note $1+\xi^2=2Mr/\Sigma$. We also use $\nabla_aF_{bc}=W_{bcad}\xi^d$ and $F_{ab}=\tfrac14W_{abcd}k^{cd}$. We use the rank-two Ricci identity in this convention, \begin{equation}\label{sm:eq:ricci-comm} [\nabla_a,\nabla_b]h_{cd}=-W_{abc}{}^e h_{ed}-W_{abd}{}^e h_{ce}. \end{equation}

\subsection{The first-order symmetry 
\texorpdfstring{$\mathcal K_1$}{K1}}

The operator~\eqref{eq:calK1} combines the algebraic action $-2k_{(a}{}^ch_{b)c}$ with a derivative term built from $W_b=k^{cd}\nabla_ch_{bd}-2\xi^ch_{bc}$ and the Killing symmetry $4\mu^{-2}\mathcal L_\xi$. We establish its symmetry property directly in four dimensions. An optional string interpretation is given after the proof. \emph{Covariant form.} Expressions with explicit curvature terms reduce to~\eqref{eq:calK1} using $2\big(k_{cd}[\nabla_{(a},\nabla^c]h_{b)}{}^d+k_{(a|c|}[\nabla_d,\nabla_{b)}]h^{cd}\big)=4F_{(a}{}^ch_{b)c}$, which follows from the Ricci identity~\eqref{sm:eq:ricci-comm} and $F_{ab}=\tfrac14W_{abcd}k^{cd}$.

\emph{Symmetry.} 
We show that $\mathcal K_1$ maps solutions to solutions by proving that $E[\mathcal K_1h]=Z[\mathcal K_1h]=h[\mathcal K_1h]=0$ whenever $(E,Z,h)=0$. These conditions are equivalent to $(\mathfrak L,Y,h)=0$, since $Z=Y-\nabla h$ and $E_{ab}=\tfrac12\mathfrak L_{ab}+\nabla_{(a}v_{b)}$.
On a Ricci-flat background, the linearised Ricci tensor annihilates an infinitesimal coordinate variation, so $\delta R[\nabla_{(a}W_{b)}]=0$. This is a property of the linearised Ricci operator; the mass term is retained when checking the full Fierz--Pauli equations. The remaining step uses a $\mu$-independent identity for the principal tensor $k$, which follows from \eqref{sm:eq:I} and the Ricci identity~\eqref{sm:eq:ricci-comm}:
\begin{equation}
-2\,\delta R[k\cdot h]_{ab}+\nabla_{(a}W_{b)}=-2k_{(a}{}^c\delta R_{b)c}+k_{(a}{}^c\nabla_{|c|}Z_{b)}+\xi_{(a}Z_{b)}+g_{ab}\,\xi^c(Z_c+\nabla_ch).
\end{equation}
Combining these identities gives, for every symmetric $h$,
\begin{align*}
h[\mathcal K_1h]&=\mu^{-2}\big(2k^{cd}\nabla_cZ_d-6\,\xi^cZ_c\big),\\
Z[\mathcal K_1h]_a&=\mu^{-2}\big(8\xi^cE_{ac}+\xi^c\nabla_cZ_a-2k^{cd}\nabla_cE_{ad}+F_a{}^cZ_c\big)-\xi_a\,h-k_a{}^c\big(Z_c+\nabla_ch\big),\\
E[\mathcal K_1h]_{ab}&=\mu^{-2}\big(4\xi^c\nabla_cE_{ab}+8F_{(a}{}^cE_{b)c}-2g_{ab}\,\xi^c\nabla_cE+4g_{ab}\,\xi^c\nabla^dE_{cd}\big)-2k_{(a}{}^cE_{b)c}+k_{(a}{}^c\nabla_{|c|}Z_{b)}+\xi_{(a}Z_{b)}-g_{ab}\,\xi^cZ_c ,
\end{align*}
with $E=E^c{}_c$. The right-hand sides vanish when $(E,Z,h)=0$, so $\mathcal K_1$ maps solutions of~\eqref{eq:FP} to solutions.

\emph{Intertwining.} For the $s=2$ row of~\eqref{eq:intertwine}, we start from $\Psi_0[\mathcal K_1h]=\delta C_{lmlm}[\mathcal K_1h]$, with $\mathcal K_1$ given by~\eqref{eq:calK1}.
The derivative term drops out because $\Psi_0$ is invariant under $h\to h+\nabla_{(a}W_{b)}$ on the type-D background~\cite{Teukolsky:1972my,Teukolsky:1973ha}. Also, $\Psi_0[\mathcal L_\xi h]=\mathcal L_\xi\Psi_0[h]$ because the Kinnersley tetrad is invariant under $\xi$. The relevant components of $-2k\cdot h$ are $(\zeta+\bar\zeta)h_{ll}$, $\zeta h_{lm}$ and $(\zeta-\bar\zeta)h_{mm}$. Since $\eth(\zeta+\bar\zeta)=0$ and $\thorn(\zeta-\bar\zeta)=0$, the formula for $\Psi_0$ reduces to
\begin{equation}
\Psi_0[\mathcal K_1h]=(\zeta-\bar\zeta)\Psi_0+\bar\zeta(\eth-\bar\tau')X_1+\tfrac4{\mu^2}\mathcal L_\xi\Psi_0=d_2\Psi_0+\tfrac1{\sqrt2}L^\dagger_{-1}X_1,
\end{equation}
which is $\tfrac12[K_1\vO[h]]_{s=2}$ for every $h$.
For the $s=1,0$ rows we apply $X_1[\cdot]=F_{lml}[\cdot]$ and $h_{ll}[\cdot]$ to~\eqref{eq:calK1}. Here the derivative term contributes, and expanding $W$ gives
\begin{align*}
X_1[\mathcal K_1h]&=\tfrac{2}{\mu^2}\zeta(\eth'-4\tau')\Psi_0+d_1X_1-\tfrac32\bar\zeta(\eth-2\bar\tau')h_{ll}+\tfrac{1}{\mu^2}\Big[4R_{lm}+\big(\tfrac{3\bar\tau'}{\bar\rho}-\tfrac{\bar\tau'}{\rho}\big)R_{ll}+\tfrac1\rho\eth R_{ll}+\tfrac{2\bar\rho}{\rho}R_{lm}-\big(\tfrac1\rho+\tfrac1{\bar\rho}\big)\thorn R_{lm}\Big],\\
h_{ll}[\mathcal K_1h]&=-\tfrac{2}{\mu^2}\big[\zeta(\eth'-3\tau')X_1+\bar\zeta(\eth-3\bar\tau')\bar X_1\big]+\tfrac{1}{\mu^2}\big[-2Z_l+4r\,R_{ll}-2r\,\thorn Z_l\big].
\end{align*}
These are $[K_1\vO[h]]_{s=1}/\sqrt2$ and $[K_1\vO[h]]_{s=0}$, respectively, plus sources that vanish on shell.
Together these identities establish $\vO[\mathcal K_1h]=K_1\vO[h]$ on solutions, proving~\eqref{eq:intertwine} for $i=1$.\par\medskip\noindent \emph{Optional string interpretation.} On the Kerr string, the momentum $\mu$ along $z$ is the four-dimensional mass. The form $\hat Y=k\wedge dz$ is a closed conformal Killing-Yano 3-form. Its operator on symmetric tensors is \begin{equation} (\mathcal OH)_{AB}=\hat Y_{(A}{}^{CD}\hat\nabla_{|C|}H_{B)D}+2(\xi\wedge dz)_{(A}{}^CH_{B)C}. \end{equation} Applied to $H=h\,e^{{\rm i}\mu z}$ with $H_{zA}=0$, it gives $(\mathcal OH)_{ab}=-{\rm i}\mu k_{(a}{}^ch_{b)c}$, $(\mathcal OH)_{az}=W_a/2$ and $(\mathcal OH)_{zz}=0$, with the common Fourier factor suppressed. The compensating coordinate variation generated by $\hat X_a=-W_a/({\rm i}\mu)$ removes the $az$ block and adds ${\rm i}\mu^{-1}\nabla_{(a}W_{b)}$ to the $ab$ block. Dividing the resulting $ab$ block by ${\rm i}\mu/2$ and adding $4\mu^{-2}\mathcal L_\xi h$ gives $\mathcal K_1h$ in~\eqref{eq:calK1}. 

\subsection{The second-order symmetry \texorpdfstring{$\mathcal K_2$}{K2}}
\emph{Symmetry.}
We show that $\mathcal K_2$ preserves the field equation, divergence and trace constraints in~\eqref{eq:FP-Leom}. The following four identities follow from~\eqref{sm:eq:I} and the Ricci identity~\eqref{sm:eq:ricci-comm}. Let $\mathcal C=\nabla_c(K^{cd}\nabla_d\,\cdot)$ act componentwise and $\mathcal K_v=\mathcal C-\mathcal L_{\rm sh}$.
\begin{itemize}
\item[(A)] \emph{Trace.} $h[\mathcal K_2h]=\mathcal K_vh+6\xi^2h-4\chi^cY_c$. To prove it, trace $\mathcal K_2$ term by term. The curvature term traces to $2f^{bc}f^{de}(W_{cdb}{}^kh_{ke}+W_{cde}{}^kh_{bk})$, which vanishes by antisymmetry of $f$. The two $\xi k\nabla h$ terms cancel, $4\xi^ck_{cd}\nabla_bh^{bd}=-4\chi^dY_d$, and the algebraic terms leave $6\xi^2h$.
\item[(B)] \emph{Divergence.} $Y[\mathcal K_2h]_a=\mathcal K_vY_a+(7\xi^2-3)Y_a-4\xi_a\xi^cY_c+2k_a{}^c\xi^d\nabla_cY_d-4k_a{}^c\xi^d\nabla_dY_c+2\xi_ak^{cd}\nabla_cY_d-2\chi^d(\nabla_aY_d+\nabla_dY_a)+2\xi^2\nabla_ah-2\xi_a\xi^c\nabla_ch+6F_{ac}\xi^ch$.
\item[(C)] \emph{Pure gauge.} $\mathcal K_2[\nabla_{(a}V_{b)}]=\nabla_{(a}(\mathcal K_vV)_{b)}+\mathcal G[V]_{ab}$, with $\mathcal G$ an explicit second-order operator in $V$.
\item[(D)] \emph{Lichnerowicz.} With $\Delta_Lh_{ab}=-\Box h_{ab}+2W_{acbd}h^{cd}$, 
\begin{align*}
\tfrac12[\Delta_L,\mathcal K_2]h_{ab}&=4\xi_{(a}\xi^c\nabla_{|c|}Y_{b)}-4\xi^2\nabla_{(a}Y_{b)}-8F_{(a|c|}\xi^cY_{b)}-4(Fk)_{(a}{}^d\nabla_{|d|}Y_{b)}-2\big(W_{acbd}+W_{bcad}\big)\chi^cY^d\\
&\quad-4F_{(a}{}^c\xi_{b)}\nabla_ch+4g_{ab}F^{cd}\xi_c\nabla_dh-2W_{acbd}\xi^c\xi^dh+2F_{(a}{}^cF_{b)c}h-2g_{ab}F^{cd}F_{cd}h .
\end{align*}
\end{itemize}
The commutator in (D) is at most second order for structural reasons. The principal symbol of $\Delta_L$ is $|p|^2\,\mathrm{Id}$, and that of $\mathcal K_2$ is $-K^{cd}p_cp_d\,\mathrm{Id}$ plus the $\mathcal L_{\rm sh}$ term. The third-order symbol is the Poisson bracket $\{g^{ab}p_ap_b,K^{cd}p_cp_d\}$, which vanishes because $K$ is a Killing tensor and $\xi,\eta$ are Killing vectors. The content of (D) is that the first-order and curvature terms of $\mathcal K_2$ make the remainder depend on $h$ only through the de Donder data $(Y,h)$.

On a solution, $Y=0$, $h=0$ and $(\Delta_L+\mu^2)h=0$. Then (A) and (B) give $Y[\mathcal K_2h]=h[\mathcal K_2h]=0$, and (D) gives $(\Delta_L+\mu^2)\mathcal K_2h=0$. Thus $\mathcal K_2$ maps solutions of the four-dimensional Fierz--Pauli equations~\eqref{eq:FP} to solutions.

\emph{Intertwining.} To prove~\eqref{eq:intertwine} for $i=2$, we evaluate $\Psi_0[\mathcal K_2h]=\delta C_{lmlm}[\mathcal K_2h]$, $X_1[\mathcal K_2h]=F_{lml}[\mathcal K_2h]$ and $h_{ll}[\mathcal K_2h]$, using~\eqref{eq:field-components}.
Computing the tetrad components of $\mathcal K_2h$ with the connection rules and the commutators gives, for every $h$,
\begin{align*}
\Psi_0[\mathcal K_2h]&=K_{2,22}\Psi_0+2\mu^2\Sigma\tau X_1+\mathcal J_2,\\
X_1[\mathcal K_2h]&=K_{2,11}X_1-4\Sigma\tau\Psi_0-3\mu^2\Sigma\tau h_{ll}+\mathcal J_1,\\
h_{ll}[\mathcal K_2h]&=K_{2,00}h_{ll}+4\Sigma\tau(X_1-\bar X_1)+\mathcal J_0,
\end{align*}
where $\mathcal J_s$ are linear in $R=(E,Z,h)$ and its derivatives. With $c=a\cos\theta$ and $\hat Y_0\equiv-2R_{ll}+2\thorn Z_l+\thorn^2h$ (the operator $Y_0$ of~\eqref{sm:eq:thm} without spin coefficients), the $s=0$ residual is a commutator with Carter's tensor,
\begin{equation}
\mathcal J_0=r^2\,\hat Y_0[R]-\hat Y_0[\Sigma R]=2c^2\big(R_{ll}-\thorn Z_l\big)-c^2\thorn^2h-4r\big(Z_l+\thorn h\big)-2h .
\end{equation}
For $s=1,2$ the residuals have the same leading part, $-c^2Y_s[R]$, plus lower-order terms in $R$, $Z$ and $h$ with coefficients built from $r$, $c$ and the spin coefficients. Since $\mathcal J_s$ vanish on shell, $\vO[\mathcal K_2h]=K_2\vO[h]$ on solutions. This proves~\eqref{eq:intertwine} for $i=2$.

\subsection{Commutativity and the spectral curve}
A direct computation with the coordinate forms in Sec.~\ref{sm:operators} gives $[K_1,K_2]=0$, identically in $a,\omega,m,\mu$ (the existence of joint eigenfunctions $\bm S$ alone would not establish this operator identity). We next establish the operator relation $F(K_1,K_2)=0$ in~\eqref{eq:FK} by a finite-dimensional calculation.

Write $K_1=B\,\partial_\theta+C$ and $K_2=\partial_\theta^2+\cot\theta\,\partial_\theta+V$, where $B,C,V$ are $5\times5$ matrices of functions of $\theta$. The local solutions of $(K_2-\nu)\bm S=0$ form a ten-dimensional space $\mathcal V_\nu$, parametrised by $(\bm S,\bm S')$ at any $\theta_0\in(0,\pi)$.
Since $[K_1,K_2]=0$, $K_1$ maps $\mathcal V_\nu$ into itself. Its action on the initial data is
\begin{equation}\label{eq:M}
\begin{pmatrix} 
(K_1 \bm{S})\\
(K_1 \bm{S})'
\end{pmatrix}_{\theta_0}
=\mathsf M(\nu) \begin{pmatrix}
\bm{S}\\
\bm{S}'
\end{pmatrix}_{\theta_0}
% The joint-eigenfunction condition is stated below.
\quad \hbox{with}\quad
\mathsf M(\nu)=
\begin{pmatrix} C & B\\ C'-BV+\nu B & B'+C-\cot\theta\,B
\end{pmatrix}_{\theta_0},
\end{equation}
Here we used $\bm S''=\nu\bm S-\cot\theta\,\bm S'-V\bm S$ to eliminate $\bm S''$. For a joint eigenfunction, the initial-data vector is also an eigenvector of $\mathsf M(\nu)$ with eigenvalue $\lambda$.
Different $\theta_0$ give conjugate matrices, since they represent the same map in different bases. At $\theta_0=\pi/2$ the entries are polynomial in $a,m,\omega,\nu$ and Laurent polynomial in $\mu^2$. 
Evaluating the matrix at $\theta_0=\pi/2$ and computing its characteristic polynomial and the matrix polynomial separately gives
\begin{equation}
\det\big(\lambda-\mathsf M(\nu)\big)=\Big(\tfrac{F(\lambda,\nu)}{F_5}\Big)^2,\qquad F\big(\mathsf M(\nu),\nu\big)=0,
\end{equation}
where $F=\sum_{k=0}^5F_k\lambda^k$ is given in~\eqref{sm:eq:Fexpl} and $F_5=-\tfrac43\mu^6$. Thus the characteristic polynomial is the square of the monic quintic $F/F_5$. When its five roots are distinct, $F(\cdot,\nu)/F_5$ is the minimal polynomial of $K_1$ on $\mathcal V_\nu$, and each eigenspace has dimension two. At coincident roots the matrix identity still holds, but the minimal polynomial and eigenspace structure require separate examination.

It follows that the matrix differential operator $R=F(K_1,K_2)$ annihilates $\mathcal V_\nu$ for every $\nu$. On $\mathcal V_\nu$, iterating $S''=\nu S+\dots$ gives $S^{(2n)}=\nu^nS+O(\nu^{n-1})$ and $S^{(2n+1)}=\nu^n(S'+\gamma_nS)+O(\nu^{n-1})$ at $\theta_0$, with $\gamma_n$ independent of $\nu$. Write $R=\sum_{k\le 2n+1}r_k\partial_\theta^k$. The coefficient of $\nu^n$ in $RS|_{\theta_0}=0$ is $r_{2n}S+r_{2n+1}(S'+\gamma_nS)$. Since $S(\theta_0)$ and $S'(\theta_0)$ are arbitrary, $r_{2n+1}(\theta_0)=r_{2n}(\theta_0)=0$. Descending in $n$, every $r_k$ vanishes at every $\theta_0$, so $R=0$. Applying $F(K_1,K_2)=0$ to a joint eigenfunction then gives $F(\lambda,\nu)=0$, as in~\eqref{eq:F}.

\section{Radial reduction}\label{sm:s:radial}

On a joint eigenspace~(6) %\eqref{eq:sep},
the transport identity~(9)%~\eqref{eq:transport} 
becomes the third-order equation $L_3f=0$, Eq.~(11). %~\eqref{eq:L3}. 
At large $r$ its solutions behave as $e^{\kappa r}r^p$ with $(\kappa-{\rm i}\omega)(\kappa^2-\mu^2+\omega^2)=0$. The branch $\kappa={\rm i}\omega$ has massless asymptotics. It is introduced by differentiating $(T_s)$ in~\eqref{eq:Ts} and using $(B_s)$ in~\eqref{eq:Bs} to eliminate $G_s$, giving~\eqref{eq:transport}. The physical first integral below excludes this extra branch. The exponential $\exp({\rm i}\!\int\!K/\Delta\,dr)$ is the homogeneous solution for $L_2f$, rather than, in general, an exact solution for $f$.

\paragraph{The first integral.} Let $\mathcal A[A]$ denote the left-hand side of the adjoint equation~\eqref{eq:Aeq}. For any $A(r)$, a direct computation gives
\begin{equation*}
D\circ L_2=A\,L_3+\mathcal A[A],
\end{equation*}
with $L_2$ defined in~\eqref{eq:radial} and $L_3$ in~\eqref{eq:L3}.
The large-$r$ balance of the $A'$ and $A$ terms in $\mathcal A[A]$ is $\mu^2(4-\deg A)\,r^{\deg A}$, so a polynomial solution is a quartic. With $A=\sum_{k=0}^4A_kr^k$ and $A_4=\mu^4$, setting the coefficients of $r^4,\dots,r^1$ in $\mathcal A[A]$ to zero determines $A_3,\dots,A_0$:
\begin{align*}
A_3&=2\mu^2(\lambda\mu^2+4{\rm i}\omega),\qquad
A_2=2\lambda^2\mu^4+10{\rm i}\lambda\mu^2\omega-2\mu^2\nu-4\mu^2-8\omega^2+2a^2\mu^2\omega^2-4am\mu^2\omega,\\
A_1&=\tfrac43\lambda^3\mu^4+\tfrac{20}3{\rm i}\lambda^2\mu^2\omega+\lambda\big(\tfrac{10}3a^2\mu^2\omega^2-\tfrac{20}3am\mu^2\omega-\tfrac{10}3\mu^2\nu-\tfrac83\mu^2-\tfrac{16}3\omega^2\big)\\
&\quad+8M\mu^2+8{\rm i}a^2\mu^2\omega+8{\rm i}a^2\omega^3-8{\rm i}am\mu^2-16{\rm i}am\omega^2-8{\rm i}\nu\omega,\\
A_0&=\tfrac23\lambda^4\mu^4+4{\rm i}\lambda^3\mu^2\omega+\lambda^2\big(\tfrac83a^2\mu^2\omega^2-\tfrac{16}3am\mu^2\omega-\tfrac83\mu^2\nu-\tfrac43\mu^2-6\omega^2\big)\\
&\quad+\lambda\big(4M\mu^2+10{\rm i}a^2\mu^2\omega+\tfrac{32}3{\rm i}a^2\omega^3-10{\rm i}am\mu^2-\tfrac{64}3{\rm i}am\omega^2-\tfrac{32}3{\rm i}\nu\omega-\tfrac43{\rm i}\omega-\tfrac{8{\rm i}\omega^3}{3\mu^2}\big)\\
&\quad+\nu^2+\nu\big(2-2a^2\omega^2+4am\omega+\tfrac{8\omega^2}{\mu^2}\big)+4{\rm i}M\omega+(a\omega)^2(a\omega-2m)^2-6a^2\mu^2-30a^2\omega^2-\tfrac{8a^2\omega^4}{\mu^2}+32am\omega+\tfrac{16am\omega^3}{\mu^2}.
\end{align*}
The remaining constant coefficient is $\mathcal A[A]=F(\lambda,\nu)$, which is independent of $M$:
\begin{align}\label{sm:eq:Fexpl}
%F
F(\lambda,\nu)&=-\tfrac43\mu^6\lambda^5-\tfrac{40}3{\rm i}\mu^4\omega\lambda^4+\lambda^3\big(\tfrac{20}3\mu^4\nu+\tfrac{16}3\mu^4+44\mu^2\omega^2-\tfrac{20}3a^2\mu^4\omega^2+\tfrac{40}3am\mu^4\omega\big)\nonumber\\
&\quad+\lambda^2\big(\tfrac{148}3{\rm i}\mu^2\nu\omega+\tfrac{80}3{\rm i}\mu^2\omega+\tfrac{160}3{\rm i}\omega^3-28{\rm i}a^2\mu^4\omega-\tfrac{148}3{\rm i}a^2\mu^2\omega^3+28{\rm i}am\mu^4+\tfrac{296}3{\rm i}am\mu^2\omega^2\big)\nonumber\\
&\quad+\lambda\Big(-\tfrac{16}3\mu^2\nu^2-\tfrac{40}3\mu^2\nu-\tfrac{16}3\mu^2-\tfrac{320}3\nu\omega^2-\tfrac{64}3\omega^2-\tfrac{64\omega^4}{3\mu^2}+24a^2\mu^4+\tfrac{32}3a^2\mu^2\nu\omega^2+\tfrac{568}3a^2\mu^2\omega^2+\tfrac{320}3a^2\omega^4\nonumber\\
&\qquad-\tfrac{64}3am\mu^2\nu\omega-\tfrac{608}3am\mu^2\omega-\tfrac{640}3am\omega^3-\tfrac{16}3a^2\mu^2\omega^2(a\omega-2m)^2\Big)\nonumber\\
&\quad-16{\rm i}\nu^2\omega+\nu\big(-32{\rm i}\omega-\tfrac{64{\rm i}\omega^3}{\mu^2}+16{\rm i}a^2\mu^2\omega+32{\rm i}a^2\omega^3-16{\rm i}am\mu^2-64{\rm i}am\omega^2\big)\nonumber\\
&\quad-16{\rm i}a^4\omega^3(\mu^2+\omega^2)+16{\rm i}am\,a^2\omega^2(3\mu^2+4\omega^2)-32{\rm i}a^2m^2\omega(\mu^2+2\omega^2)\nonumber\\
&\quad+128{\rm i}a^2\mu^2\omega+288{\rm i}a^2\omega^3+\tfrac{64{\rm i}a^2\omega^5}{\mu^2}-32{\rm i}am\mu^2-320{\rm i}am\omega^2-\tfrac{128{\rm i}am\omega^4}{\mu^2}.
\end{align}
By $F(K_1,K_2)=0$, $F(\lambda,\nu)$ vanishes on every joint eigenspace. Hence $A$ is a quartic solution of the adjoint equation and \begin{equation}\label{eq:DL2-AL3} D\circ L_2=A\,L_3. \end{equation} The polynomial in~\eqref{sm:eq:Fexpl} is $F(\lambda,\nu;a,m,\omega,\mu)$ in~\eqref{eq:F} of the Letter.

\paragraph{The physical relation.} Let $f$ solve~(11). %\eqref{eq:L3}. 
Then $D(L_2f)=0$, i.e.\ $L_2f=C\exp({\rm i}\!\int\!K/\Delta\,dr)$, which carries the massless asymptotic branch discussed above. For Fierz--Pauli fields $C=0$, so that~(13) %\eqref{eq:radial} 
holds. This follows from the local identity $\mathcal L_2[\vO[h]]=0$ of Sec.~\ref{sm:s:recon}. It is also seen from the asymptotics: each component of a Fierz--Pauli field behaves at large $r$ as $e^{\pm\sqrt{\mu^2-\omega^2}\,r}r^p$, and so does $L_2f$, which is incompatible with $e^{{\rm i}\omega r}r^q$ unless $C=0$. Equivalently, $L_2f$ is the bilinear concomitant of $f$ with the adjoint solution $A\exp(-{\rm i}\!\int\!K/\Delta\,dr)$, which is dual to the massless branch.

\paragraph{Boundary behaviour.} Eq.~(13)  %~\eqref{eq:radial} 
has regular singular points at $r_\pm$, the apparent singularities at the zeros of $A$, and an irregular singular point at infinity. At the outer horizon the ingoing solution behaves as $f\sim(r-r_+)^{-2-{\rm i}K_+/\Delta'_+}$, with $K_+=K(r_+)$ and $\Delta'_+=r_+-r_-$, as for the spin-2 Teukolsky function. At infinity $f\sim e^{\pm\sqrt{\mu^2-\omega^2}\,r}r^p$.

%%%%%%%%%%%%%%%%%%%%
\section{Reconstruction and the converse}\label{sm:s:recon}

We describe the reconstruction of $h_{ab}$ from the five GHP components $\vO$ in~\eqref{eq:Omega}. The joint angular equations~\eqref{eq:sep}, the spectral relation~\eqref{eq:F} and the physical second-order radial equation~\eqref{eq:radial} define the separated problem. Before separation, the system $\mathfrak S$ used below consists of the transport identity~\eqref{eq:transport} together with the first integral~\eqref{sm:eq:L2op}.

\paragraph{The separated system.} Fix $(\omega,m)$. Let $\mathfrak S$ be the system formed by the transport identity~(9)  %~\eqref{eq:transport}
and by the first integral
\begin{equation}\label{sm:eq:L2op}
\mathcal L_2[\vO]\equiv A(r;K_1,K_2)(\Dr+K_2)\vO-\Delta A'(r;K_1,K_2)\,\partial_r\vO\\
+\big[\Delta A''-(2\Delta'+{\rm i}K)A'\big](r;K_1,K_2)\,\vO=0,
\end{equation}
This is the operator form of~\eqref{eq:radial}, with $\lambda,\nu$ replaced by the commuting operators $K_1,K_2$. On a separated joint eigenfunction $\vO=f\bm S$, $\mathcal L_2[\vO]=\bm S\,L_2f$, so every solution of the physical radial equation satisfies the first integral. Moreover, $D\circ L_2=A L_3$ on the spectral curve, by~\eqref{eq:DL2-AL3}; hence $A L_3f=0$ and $L_3f=0$ wherever $A\neq0$, extending across its isolated zeros for smooth solutions. Thus $\mathcal E[\vO]=\bm S\,L_3f=0$ as well.

\paragraph{Theorem.} For every $(\omega,m)$, the following hold on any open set:
\begin{enumerate}
\item[(i)] if $h$ solves the Fierz--Pauli equations, then $\vO[h]$ solves $\mathfrak S$;
\item[(ii)] if $h$ solves the Fierz--Pauli equations, then $h=\mathcal Q[\vO[h]]$;
\item[(iii)] if $\vO$ solves $\mathfrak S$, then $h=\mathcal Q[\vO]$ solves the Fierz--Pauli equations and $\vO[h]=\vO$.
\end{enumerate}
Thus $h\mapsto\vO[h]$ is a bijection from Fierz--Pauli modes onto solutions of $\mathfrak S$, with inverse $\mathcal Q$. In particular every separated solution defines the Fierz--Pauli field $h=\mathcal Q[f\bm S]$, and by the completeness result below every Fierz--Pauli mode is a convergent sum of such fields, with finite Jordan-chain combinations included when necessary at complex frequencies. Part (i) also shows that $L_2f=0$ holds identically, not only for solutions with massive asymptotics.

\paragraph{Completeness.} Angular completeness at fixed $(\omega,m)$ follows by treating $K_2$ as a bounded perturbation of the spin-weighted spherical operator~\cite{Markus,Wyss}. For real $\omega$, a positive weighted inner product makes $K_2$ self-adjoint and $K_1$ skew-adjoint, giving a complete orthonormal basis of joint eigenfunctions. For complex $\omega$, the expansion is instead organised into finite-dimensional joint root-space blocks, allowing Jordan chains and their associated triangular radial systems. For smooth fields regular on the axis, reconstruction with $\mathcal Q$ gives the corresponding convergent expansion of Fierz--Pauli fields. This establishes angular completeness; completeness of quasinormal or quasibound modes in time is not addressed.

\paragraph{The reconstruction operator $\mathcal Q$.} It reads
\begin{equation}\label{sm:eq:Qrec}
h_{ab}=\mathcal Q_{ab}[\vO]\equiv\sum_{f,w}c^{ab}_{f,w}\;w\,\vO_f ,
\end{equation}
summed over $\vO_f\in\{\Psi_0,X_1,h_{ll},\bar X_1,\bar\Psi_0\}$ and over words $w$ in $\thorn'$, $\eth$, $\eth'$ (letters act from right to left). The words have length at most six on $\Psi_0,\bar\Psi_0$ and at most four on the other components. The coefficients are polynomials in $\tau$, $\Psi_2$ and $\mathcal L_\xi$, and Laurent polynomials in $\zeta$, $\bar\zeta$, $\rho'$ and $\mu^2$, with rational coefficients; $\tau$ occurs at most linearly. There are 579 nonzero coefficients, including the trivial coefficient for $h_{ll}$; conjugate components are counted separately. The complete list is given in the ancillary file \texttt{Q\_reconstruction.wl}, as entries $\{$component, field, word, coefficient$\}$; its header states the coordinate, tetrad, GHP and normalisation conventions used here. For example, $h_{ll}$ is reproduced trivially, and the $\eth'X_1$ and $\thorn'\thorn'X_1$ terms of $h_{lm}$ are $-2\mu^{-2}\tau\zeta\,\eth'X_1$ and $\tfrac23\mu^{-2}\rho'^{-1}\,\thorn'\thorn'X_1$. Since $\mathcal L_\eta$ does not occur and no coefficient has a pole in $\omega$, $\mathcal Q$ is the same operator for every $m$ and every frequency $\omega$. Each coefficient carries a definite GHP boost and spin weight.

\paragraph{Proof.} Each statement (i)-(iii) says that some expression vanishes on the solutions of a system. We prove it by reducing the expression to zero with respect to a passive form of that system. By \eqref{sm:eq:Lxi}, on modes $\thorn$ is a combination of $\thorn'$, $\eth$, $\eth'$ and $\mathcal L_\xi$, and we eliminate it throughout. The $\phi$-dependence is fixed by the first-order equation $\mathcal L_V=-{\rm i}K$ for $V=(r^2+a^2)\xi+a\eta=\Sigma(\bar\zeta\bar\rho'\,l-\zeta\rho\,n)$. Derivatives are ordered by total order, then by the number of $\thorn'$, then by the number of $\eth$. All coefficients are kept as weighted GHP quantities, so that derivatives of coefficients are covariant. The relations particular to the Kinnersley frame ($\zeta\rho=-1$, $\bar\tau=-\tau$, $a^2=\Delta-r^2+2Mr$) are imposed only on the final remainders.

\emph{Fierz--Pauli metric equations.} Order the components as $h_{m\bar m},h_{nn},h_{nm},h_{n\bar m},h_{ln},h_{ll},h_{lm},h_{l\bar m},h_{mm},h_{\bar m\bar m}$. The equations $h=0$, $Z_a=0$, $E_{ab}=0$ are then equivalent to fourteen equations with leading terms
\begin{gather*}
h_{m\bar m};\qquad \thorn'h_{ln},\ \thorn'h_{nn},\ \thorn'h_{nm},\ \thorn'h_{n\bar m};\qquad
\thorn'^2h_{ll},\ \thorn'^2h_{lm},\ \thorn'^2h_{l\bar m},\ \thorn'^2h_{mm},\ \thorn'^2h_{\bar m\bar m};\\
\thorn'\eth\,h_{ll},\ \thorn'\eth\,h_{lm},\ \thorn'\eth\,h_{l\bar m},\ \eth^2h_{n\bar m}.
\end{gather*}
The first is the trace and the next four are the divergence. The next five are $E_{ll}$, $E_{lm}$, $E_{l\bar m}$, $E_{mm}$ and $E_{\bar m\bar m}$. The last four are $E_{ln}$, $E_{nm}$, $E_{n\bar m}$ and $E_{nn}$, reduced by the trace and the divergence; $E_{m\bar m}$ is then a consequence. Every integrability condition between two of these equations reduces to zero, so they form a passive (Gr\"obner) basis.

\emph{The separated system.} Write $u_s=\vO_s/c_s$. The Fourier-mode conditions $\mathcal L_\xi=-{\rm i}\omega$ and $\mathcal L_\eta={\rm i}m$, the five rows of $\mathcal E$ in~\eqref{eq:transport} and the five rows of $\mathcal L_2$ in~\eqref{sm:eq:L2op} have leading terms $\eth u_s$, $\thorn'^3u_s$ and $\thorn'^2\eth'^4u_s$, one of each for every component. Their leading coefficients are products of $\zeta$, $\bar\zeta$, $\rho$, $\rho'$, $\mu^2$, $\tau$ and $\bar\tau$; in particular $A(r)$ never has to be inverted. Every integrability condition reduces to zero, so these fifteen equations are passive. Like the Fierz--Pauli system, at each order above the sixth they leave ten derivatives free: two for each degree of freedom.

\emph{The identities.} (i) $\mathcal E[\vO[h]]$ and $\mathcal L_2[\vO[h]]$ reduce to zero with respect to the Fierz--Pauli basis and the Fourier-mode conditions. (ii) $\mathcal Q[\vO[h]]-h$ reduces to zero with respect to the Fierz--Pauli basis, for each of the ten components. (iii) For $h=\mathcal Q[\vO]$, the trace vanishes identically, and $Z_a[h]$, $E_{ab}[h]$ and $\vO[h]-\vO$ reduce to zero with respect to the basis of $\mathfrak S$. The first integral \eqref{sm:eq:L2op} is essential here. With the transport identity~\eqref{eq:transport} alone, $E_{nn}$, $E_{nm}$, $E_{n\bar m}$, $E_{mm}$ and $E_{\bar m\bar m}$ do not reduce to zero, because the massless branch is not a Fierz--Pauli field. The reductions divide only by leading coefficients, so each identity holds where $\Delta\sin\theta\neq0$, and hence everywhere by continuity.

\paragraph{Regularity.} Each coefficient of $\mathcal Q$ of spin weight $s$ has the form $\sin^{|s|}\theta\,G(r,\cos\theta)$, with $G$ rational and regular at $\cos\theta=\pm1$ for $r>r_+$. So $\mathcal Q$ maps fields regular on the axis to fields regular on the axis. At the outer horizon we use ingoing coordinates $(v,r,\theta,\tilde\phi)$, with $dv=dt+(r^2+a^2)dr/\Delta$ and $d\tilde\phi=d\phi+a\,dr/\Delta$, and the Hartle-Hawking tetrad $l_H=(\Delta/2\Sigma)\,l$, $n_H=(2\Sigma/\Delta)\,n$, $m$, which is smooth across the future horizon. A quantity of boost weight $b$ is multiplied by $(\Delta/2\Sigma)^b$ in this tetrad. The radial function $f$ is ingoing when $(\Delta/2\Sigma)^2\vO\,e^{{\rm i}(\omega r_*-m\tilde r)}$ is analytic at $r_+$, where $dr_*/dr=(r^2+a^2)/\Delta$ and $d\tilde r/dr=a/\Delta$. Each coefficient of $\mathcal Q$ carries a definite GHP weight and contains at most one factor $\rho'^{-1}\propto\Delta^{-1}$, so the Hartle-Hawking components of $h=\mathcal Q[f\bm S]$ obey $h_{AB}=O(\Delta^{d_{AB}})$, with
\begin{equation*}
d_{ll}=0,\quad d_{lm}=d_{l\bar m}=-1,\quad d_{ln}=d_{mm}=d_{m\bar m}=d_{\bar m\bar m}=-2,\quad d_{nm}=d_{n\bar m}=-3,\quad d_{nn}=-4.
\end{equation*}
Write $h_{AB}=\sum_{n\ge-4}\Delta^nH^{(n)}_{AB}(\theta)$. At the lowest power, the equations $\mathfrak Lh=0$, $Z_a=0$ and $h=0$ give a linear algebraic system for $H^{(n)}$, with coefficients evaluated at $r_+$. Let $k=\omega-m\Omega_H$ and let $\kappa_+=(r_+-r_-)/[2(r_+^2+a^2)]$ be the surface gravity. The system has a nonzero solution only if $k=-{\rm i}j\kappa_+$ with $j\in\{n,n+1,n+2\}$. Restricted to the components allowed by the bounds $d_{AB}$, it has a nonzero solution only for $j\in\{-2,-1,0,1\}$. Inducting from $n=-4$ to $n=-1$ then shows that $\mathcal Q[f\bm S]$ with $f$ ingoing is regular on the future horizon, except possibly at
\begin{equation}
k=\omega-m\Omega_H\in\{-{\rm i}\kappa_+,\,0,\,{\rm i}\kappa_+,\,2{\rm i}\kappa_+\},
\end{equation}
where the argument does not conclude. For real $\omega$ only the synchronised frequency $k=0$ remains; there the two horizon exponents of $f$ also differ by an integer, so the ingoing solution itself may contain a logarithm.

\section{Schwarzschild limit}\label{sm:s:schw}
At $a=0$, $d_s=-{\rm i}s^2\omega/\mu^2$, the couplings in $K_2$ vanish, and $K_{2,ss}$ is the spin-weighted Laplacian shifted by $s^2$. On spin-weighted harmonics it equals $-\ell(\ell+1)$ for every $s$, so $\nu=-\ell(\ell+1)$. $K_1$ then acts within each $\ell$-multiplet as a $5\times5$ (or, for $\ell=1$, $3\times3$; for $\ell=0$, $1\times1$) constant matrix, and the spectral curve factorises,
\begin{equation}
F\propto\big(\lambda^2\mu^4+5{\rm i}\lambda\mu^2\omega-\mu^2\nu-2\mu^2-4\omega^2\big)\big(\lambda^3\mu^4+5{\rm i}\lambda^2\mu^2\omega-4\lambda\mu^2\nu-2\lambda\mu^2-4\lambda\omega^2-12{\rm i}\nu\omega\big).
\end{equation}
The quadratic factor carries the two axial and the cubic factor the three polar degrees of freedom. For $\ell=0$ only $h_{ll}$ is present, and $\lambda=0$; for $\ell=1$ the eigenvalues are $\lambda=-{\rm i}\omega/\mu^2$ (axial) and the roots of $\lambda^2+{\rm i}\omega\lambda/\mu^2+6/\mu^2=0$ (polar).

\end{document}